\documentstyle[12pt,epsf]{article}

\def\be{\begin{equation}}
\def\ee{\end{equation}}
\def\ba{\begin{eqnarray}}
\def\ea{\end{eqnarray}}

\def\esp#1{{\rm e}^{#1}}
\def\ii{{\rm i}}
\def\tr{{\rm tr}}
\def\Tr{{\rm Tr}}
\def\diag{{\rm diag}}
\def\ket#1{|#1\rangle}
\def\bra#1{\langle #1|}
\def\braket#1#2{\langle #1 | #2 \rangle}
\newcommand{\eq}{\begin{equation}}
\newcommand{\en}{\end{equation}}
\newcommand{\eqa}{\begin{eqnarray}}
\newcommand{\ena}{\end{eqnarray}}

\newcommand{\NP}[1]{Nucl.\ Phys.\ {\bf #1}}
\newcommand{\PL}[1]{Phys.\ Lett.\ {\bf #1}}

\newcommand{\CMP}[1]{Comm.\ Math.\ Phys.\ {\bf #1}}

\newcommand{\MPL}[1]{Mod.\ Phys.\ Lett.\ {\bf #1}}
\newcommand{\IJMP}[1]{Int.\ J.\ Mod.\ Phys.\ {\bf #1}}

\def\hepth#1{{\tt hep-th/#1}}

\def\mycaptionl#1{%
\refstepcounter{figure}
\begin{center}
\hskip 1pt\vskip -0.6cm
\begin{minipage}{12cm}
\small {\bf Fig. \hskip -3pt\arabic{figure}}: {\sl #1}
\end{minipage}
\null\hskip 1pt\vskip -0.2cm
                                          \end{center}}

\begin{document}

\begin{titlepage}
\vskip0.5cm
\begin{flushright}
DFTT 56/98\\
KUL-TF-98/36\\
\end{flushright}
\vskip0.5cm
\begin{center}
{\Large\bf Matrix string states in pure 2d Yang Mills theories.}
\end{center}
\vskip 1.cm
\centerline{\bf
M. Bill\'o\footnote{e--mail: Marco.Billo@fys.kuleuven.ac.be}}
\centerline{\sl Instituut voor theoretische fysica,}
\centerline{\sl Katholieke Universiteit Lueven, B3001 Leuven, Belgium}
\vskip 0.6cm
\centerline{\bf
M. Caselle\footnote{e--mail: caselle@to.infn.it}
A. D'Adda\footnote{e--mail: dadda@to.infn.it}
 and P. Provero\footnote{e--mail:  provero@to.infn.it}}
 \centerline{\sl   Dipartimento di Fisica
 Teorica dell'Universit\`a di Torino}
 \centerline{\sl Istituto Nazionale di Fisica Nucleare, Sezione di Torino}
 \centerline{\sl via P.Giuria 1, I-10125 Torino, Italy}
 \vskip .4 cm
\vskip 0.6cm

\begin{abstract}
We quantize pure 2d Yang-Mills theory on a torus in the gauge where
the field strength is diagonal.  Because of the topological 
obstructions to a global smooth diagonalization, we find string-like states in 
the
spectrum similar to the ones introduced by various authors
in Matrix string theory. We write explicitly the partition function, 
which generalizes the one already known in the literature, and we discuss
the role  of these states in preserving modular invariance. 
Some speculations are presented about the interpretation of 
2d Yang-Mills theory as a  Matrix string theory.  
\end{abstract}
\end{titlepage}

\setcounter{footnote}{0}
\def\thefootnote{\arabic{footnote}}

\section{Introduction}
In the last few years a lot of interesting results have been obtained on two 
dimensional gauge theories like QCD2 and pure Yang-Mills theory.
Due to the invariance of 2d Yang-Mills theory under area preserving 
diffeomorphisms and its almost topological nature its partition functions 
and a number of observables have been calculated 
exactly~\cite{m75,r90,w91,w92,bt92,blth,gpss}  
on arbitrary Riemann surfaces.
In the large $N$ limit the existence of a deconfining phase transition on 
the sphere and on the cylinder has been recognized~\cite{dk,cdmp93,gm1,gm2,bg}
as a result of a condensation of instanton contributions.
Perhaps the most interesting development has been the recognition that in the 
large $N$ limit two dimensional YM theory is a string theory. In fact
the partition function of U$(N)$ Yang Mills theory on a
two dimensional Riemann surface $M_G$ of genus $G$ counts the number of 
homotopically distinct maps  from a Riemannian world-sheet $W_g$ of genus $g$ 
to $M_G$~\cite{gt}.
A new and seemingly unrelated connection between string theory and 
two dimensional gauge theories has been developed in~\cite{motl,dvv,bc}.
By combining the conjecture of Banks {\it et al.}~\cite{bfss} with the 
compactification
of an extra spatial dimension~\cite{wt} it is argued that type IIA string theory
can be identified with the large $N$ limit of two-dimensional ${\cal N}=8$ 
supersymmetric Yang-Mills theory.
In this context the eight non compact space dimensions are represented by the 
eight scalar fields $X_i$  of the ${\cal N}=8$ supermultiplet belonging 
to the adjoint representation of U$(N)$. 
In the limit $g_{YM} \rightarrow \infty$ (which is the $g_s 
\rightarrow 0$ limit for the string coupling $g_s=1/g_{YM}$) 
these eight matrix fields commute and can be simultaneously diagonalized. 
A smooth global diagonalization however is 
in general not possible because the $N$ eigenvalues can undergo a 
permutation $P$ as one goes round a non-contractable loop in the 
compactified dimension.  
As a result the spectrum  contains  states that are associated to the cycles of  
$P$ and can be identified with string states.
Supersymmetry plays a crucial role in this scheme, as it ensures the 
cancellation of the Fadeev-Popov determinants (Vandermonde determinants 
of the eigenvalues).

In this paper we show that a similar spectrum of states arises, by the same 
mechanism, in pure Yang-Mills theory on a torus. We choose the gauge in
which the field strength $F$, treated in a first order formalism as an 
independent auxiliary field, is diagonal, and analyze the sectors arising from 
non trivial permutations of its eigenvalues as one goes round the two 
independent cycles of the torus.
A fermionic symmetry between the ghost-antighost sector and the non-diagonal 
part of the gauge fields leads to the cancellation of the Vandermonde 
determinants, provided the Riemann surface on which the theory is defined 
has zero curvature. 
This limits our analysis to the torus, and leaves the problem of its 
extension to general Riemann surfaces open to speculations.
Consider now the theory on the torus as the theory of an infinite cylinder 
taken at finite temperature and denote by $P$ be the permutation of the 
eigenvalues of $F$ as we go round the compact space dimension.   
We shall find that the states that propagate along the cylinder are in 
correspondence with the decomposition of $P$ into cycles. 
More precisely they can be described as a gas of free fermions (or bosons), 
where each fermion is associated to a cycle of $P$ and is labeled  by 
two quantum numbers: the discretized momentum $n$ and the length $k$ 
of the cycle.  
The resulting partition function is therefore different from the one so far 
produced in the literature, which corresponds to the truncation to the states 
with only cycles of order $k=1$.
The states associated to non trivial permutations $P$ are described by 
holonomies whose eigenvalues are not generic: the sets of eigenvalues on 
which $P$ acts as a cyclic permutation, say of order $k$,  
are spaced like the $k$-th roots of unity. 
The truncation to $k=1$, that corresponds to the standard quantization, 
although consistent treats the compactified space and time dimensions on a 
different footing by allowing arbitrary permutations of the eigenvalues to 
occur only in the time direction, thus breaking modular invariance.
Our generalization is characterized by arbitrary commuting permutations 
along the two generators of the torus, and hence preserves modular invariance.

The plan of the paper is as follows: in Section 2 we discuss the quantization
of YM2 in the gauge where $F$ is diagonal (unitary gauge); in Section 3 we 
calculate the contribution of the new sectors, derive the partition function
on the torus and discuss the role of modular invariance; in Section 4 we 
obtain the same results by calculating the functional integral on a cylinder
and then sewing the two ends of the cylinder; in Section 5 we discuss our 
results, especially in connection with quantization in other gauges, and add a 
few concluding remarks.
\section{YM2 in the Unitary gauge}
We begin  by reviewing the main steps involved in the 
calculation of the partition function of YM2 on an arbitrary Riemann 
surface using the so called Unitary (or torus) gauge. The full details can be
found in Ref. \cite{blth}.
Let us consider the partition function
\begin{equation}
Z(\Sigma_g,t)=\int [dA][dF]\exp\left\{-\frac{t}{2}\tr \int_{\Sigma_g}
d\mu\, F^2+ \ii\, \tr \int_{\Sigma_g} f(A) F \right\}~,
\label{partriemann}
\end{equation}
where $d \mu$ is the volume form on $\Sigma_g$ and $f(A) $ is given by
\begin{equation}
f(A)= d A -\ii A\wedge A~.
\label{effea}
\end{equation}
In Eq.s (\ref{partriemann}) and (\ref{effea}) $F$ is a $N \times N$ hermitian 
matrix and $A$ is a one form on $\Sigma_g$ with values on the space of hermitian
matrices.
The usual Yang-Mills action can be recovered from (\ref{partriemann}) by 
performing the Gaussian integral over $F$.
The Unitary gauge consists in conjugating the $N\times N$ hermitian 
matrix  $F$ into a diagonal form, namely into its Cartan sub-algebra.
This can always be done, at least locally, by a gauge transformation $g$:
\begin{equation}
g^{-1} F g = \diag(\lambda)~.\label{Pgaugefix}
\end{equation}
The gauge fixed action, including the appropriate Faddeev-Popov ghost term,
can be written as the sum of two terms:
\begin{equation}
S_{\rm BRST}(\Sigma_g,t) = S_{\rm Cartan} + S_{\rm off-diag}~,
\label{action}
\end{equation}
where $S_{\rm Cartan}$ involves the diagonal part of $A_{\mu}$ and exhibits
a residual U$(1)^N$ gauge invariance:
\begin{equation}
S_{\rm Cartan}= \int_{\Sigma_g}  \sum_{i=1}^N 
\left[\frac{t}{2}\lambda_i^2 d\mu -
\ii \lambda_i d A^{(i)} \right]~,
\end{equation}
where $A^{(i)}$ is the $i$-th diagonal term of the matrix form $A$.
The Faddeev-Popov ghost term and the off-diagonal part of $A$ are contained in
$S_{\rm off-diag}$ which can be cast into the following form:
\begin{equation}
S_{\rm off-diag} =  \int_{\Sigma_g} d\mu \sum_{i>j}  (\lambda_i -
\lambda_j)\left[\hat{A}_0^{ij}\hat{ A}_1^{ji} -\hat{A}_1^{ij}\hat{A}_0^{ji} 
+ \ii ( c^{ij} \bar{c}^{ji}+ \bar{c}^{ij} c^{ji}) \right]~,
\label{offdiag}
\end{equation}
where $ \hat{A}_{a}^{ij} = E_a^{\mu} A_{\mu}^{ij} $ and $E_a^{\mu}$ denotes  
the inverse of the two dimensional vierbein.
$c^{ji}$ and $\bar{c}^{ij}$ are respectively the ghost and anti-ghost
corresponding to the gauge condition $F^{ij}=0$.
The action  (\ref{offdiag})  has some remarkable properties: it contains the 
same number of fermionic and bosonic degrees of freedom and it is symmetric, 
for each value of the composite index $[ij]$, with respect to a set of symmetry
transformation with Grassmann-odd parameters, which with abuse of language we
shall call supersymmetries. They are summarized by the following equations: 
\begin{equation}
\begin{array}{ccc}
\delta \hat{A_{0}}& =& \ii (\eta c + \zeta \bar{c})~,\cr
\delta \hat{A_{1}}& =& \ii (\xi c + \chi \bar{c})~,
\end{array}
\hskip 1cm
\begin{array}{ccc}
\delta c& =& - \chi \hat{A_{0}}+ \zeta \hat{A_{1}}~,\cr
\delta \bar{c}& =& -\xi \hat{A_{0}}+\eta  \hat{A_{1}}~,   
\end{array}
\label{supersy}
\end{equation}
where $\eta,\zeta,\xi$ and $\chi$ are the fermionic parameters and the index 
$[ij]$ has been omitted in all fields.
One would expect as a result of the supersymmetry a complete cancellation of 
the bosonic and fermionic contributions in the partition function. This is not
true in general because the supersymmetry is broken  on a generic Riemann 
surface by the measure of the functional integral.
This anomaly arises because the supersymmetric partners of the ghost anti-ghost
fields are the zero forms $\hat{A}_{a}^{ij}$, which are the component of the 
one form $A$ in the base of the vierbein. The functional integral however is
on the one form $A$, and on  a curved surface the `number' of zero forms and 
one forms does not coincide (as it is easily seen on a lattice like in Regge
calculus). The mismatch of fermionic and bosonic degrees of freedom results 
into an anomaly that has been explicitly calculated in  \cite{blth}:
\begin{equation}
\int \prod_{i>j} [dc^{ij}][d\bar{c}^{ij}][dA_{\mu}^{ij}] e^{-S_{\rm off-diag}}
= \exp \left[{1 \over 8 \pi} \int_{\Sigma_g} R \sum_{i>j} \log (\lambda_i -
\lambda_j) \right]~.
\label{anomaly}
\end{equation}
Two considerations are in order here: first that the anomaly vanishes for
surfaces with zero curvature, such as the torus or the infinite cylinder, 
second that for constant eigenvalues $\lambda_i$, the r.h.s. of (\ref{anomaly})
reduces to $\prod (\lambda_i -\lambda_j)^{2-2g}$ and it becomes divergent for
$g>1$ when two eigenvalues coincide.
We are not going to go through the whole calculation of the partition function,
which can be found elsewhere \cite{blth}; the point  is that the gauge fixing
and the following calculation of the functional integral for the U$(1)^N$ 
gauge invariant action (\ref{action}) leads to constant and integer values for 
the eigenvalues $\lambda_i$: $\lambda_i \rightarrow n_i$.
The resulting partition function of YM2 on $\Sigma_g$ for the group U$(N)$ is
then given by:
\begin{equation}
Z(\Sigma_g,t) = \sum_{\{n_i\}} {1 \over \prod_{i>j} (n_i - n_j)^{2g-2}} 
\esp{-2\pi^2 t \sum_i n_i^2}~.
\label{part} 
\end{equation}
There is nothing in the above derivation of (\ref{part}) to stop two or more 
integers $n_i$ from being coincident. On the other hand such
terms ( which we shall call ``non regular" following the terminology of Ref.
\cite{blth}) are divergent for $g>1$ and need to be regularized.
The regularization suggested in  \cite{blth} consists in adding  small mass
terms  to  $A^{ij}$.  These terms preserve the residual U$(1)^N$ gauge 
invariance but they break explicitly the supersymmetry of $S_{\rm off-diag}$.
Correspondingly the contribution to the partition function coming from 
$S_{\rm off-diag}$ is modified in the following way:
\begin{equation}
\prod_{i>j} {(n_i-n_j)^2 \over (n_i-n_j)^{2g}} \rightarrow \prod_{i>j}
{(n_i-n_j)^2 \over (n_i-n_j -m_{ij})^{2g}}~. 
\label{regularization}
\end{equation}
In (\ref{regularization}) we have kept the ghost-antighost contribution,
which is not divergent and is not affected by the regularization, separate
from the one coming from the $A^{ij}$.
Clearly after the regularization the terms with two or more coincident $n_i$'s 
vanish due to effect of the ghost contribution, while the would be divergent 
terms coming from $A^{ij}$ remain finite also for $g>1$. As a result all 
``non regular" terms are altogether suppressed.
Although not entirely satisfactory this procedure reproduces the  well known
partition function obtained both with other gauge choices and on the lattice, 
and it seems appropriate in  YM2 on Riemann surfaces with non vanishing 
curvature.
On flat surfaces however, like the torus and the infinite cylinder, the anomaly
of the fermionic symmetry (\ref{supersy}) vanishes and no regularization is
required. Hence there is no reason to add to the action terms that would break
that symmetry explicitly.
On the other hand if the supersymmetry (\ref{supersy}) is preserved nothing 
prevents non regular terms from appearing in the partition function. 
The integers $n_i$ have been interpreted on a torus (or on a cylinder) as the
discretized momenta of a gas of free fermions (or bosons \footnote{The
interpretation of the eigenvalues as bosons is associated to a quantization
which is done on the algebra rather than on the group manifold \cite{h94}.}).
Non regular terms would naturally be identified with fermions (or bosons)
carrying the same integer momentum. However  it will be
shown in the following section that a non regular term with for instance two 
coincident $n_i$'s can arise either as two states with the same momentum, 
{\it or} as one state where the two eigenvalues are exchanged as we go round a 
non contractable loop. These  states are a new feature in YM2 and they
are the exact analogue of the stringy states described by 
Dijkgraaf, E. Verlinde and H.Verlinde (DVV) in the context of Matrix string 
theory  \cite{dvv}. 
It is remarkable however that we do not need supersymmetric YM
to obtain the DVV states as  the cancellation of the Vandermonde determinants 
is ensured by the fermionic symmetry described above. 
This seems to be a peculiarity of YM2, possibly related to its interpretation as
a string theory.
\section{The partition function on the torus}
\label{sec3}
We will now concentrate on the calculation of the partition function 
(\ref{partriemann}) on the torus, defined as a square with identified opposite
sides. If we introduce a set of Euclidean coordinates $(\tau,x)$, then all
fields will obey periodic boundary conditions in both directions:
\begin{eqnarray}
\label{periodic}
A_\mu(\tau+2\pi,x)&=&A_\mu(\tau,x+2\pi)=A_\mu(\tau,x)~,\nonumber\\
F(\tau+2\pi,x)&=&F(\tau,x+2\pi)=F(\tau,x)~.
\end{eqnarray}
We have chosen for convenience to have periodicity $2\pi$ in both directions;
this is not restrictive as a rescaling of the coordinates can be absorbed in a
redefinition of the coupling $t$.
We now proceed to fix the gauge according to Eq.(\ref{Pgaugefix}). At any given
point $(\tau,x)$ the group element that conjugate the matrix $F$ into its Cartan
sub-algebra is defined up to an element of the Weyl group, namely 
in our case up to an element $P$ of the permutation group. 
So if $g(\tau,x)$ is the U$(N)$ transformation that
diagonalizes $F(\tau,x)$, any transformation $Pg(\tau,x)$ will also diagonalize
$F(\tau,x)$ to a form corresponding to a different permutation of 
the eigenvalues. If we require $g(\tau,x)$ to be continuous with its first
derivatives, then it is clear that $g(\tau,x)$ will in general be multi-valued
with boundary conditions of the type:
\begin{eqnarray}
g(\tau+2\pi,x)&=&Pg(\tau,x)~,\nonumber\\
g(\tau,x+2\pi)&=&Qg(\tau,x)~.
\end{eqnarray}
This reflects the possibility that as we go around a closed loop the eigenvalues
cross over, and undergo a permutation:
\begin{eqnarray}
\lambda_i(\tau+2\pi, x)&=&\lambda_{P(i)}(\tau,x)~,\nonumber\\
\lambda_i(\tau, x+2\pi)&=&\lambda_{Q(i)}(\tau,x)~.
\label{bcon}
\end{eqnarray}
Consistency requires that the two permutations $P$ and $Q$  commute:
\begin{equation}
PQ=QP~.
\end{equation}
After gauge fixing also the gauge field $A_\mu$ obeys generalized boundary 
conditions:
\begin{eqnarray}
\label{abc}
A_\mu(\tau+2\pi,x)&=&P^{-1}A_\mu(\tau,x)P~,\nonumber\\
A_\mu(\tau,x+2\pi)&=&Q^{-1}A_\mu(\tau,x)Q~.
\end{eqnarray}
\par
We shall give here an explicit example\footnote{Similar examples were given in 
Ref.~\cite{bt94}.} of a configuration $F(\tau,x)$ that satisfies the periodic 
boundary conditions (\ref{periodic}), but whose eigenvalues are permuted as $x 
\rightarrow x+2 \pi$. Consider for $N=2$  the following periodic configuration:
\begin{equation}
F(\tau,x)=\sin x \ \sigma_3+(1-\cos x)\  \sigma_2\label{pbcex}~,
\label{esempio}
\end{equation}
where $\sigma_i$ are the Pauli matrices. The eigenvalues are given by the 
equation
\begin{equation}
\lambda^2(\tau,x)=2(1-\cos x)=4\sin^2\frac{x}{2}
\end{equation}
namely, if we require continuity of $\lambda(\tau,x)$, by
\begin{equation}
\lambda_\pm(\tau,x)=\pm 2 \sin\frac{x}{2}~.
\end{equation}
The eigenvalues are therefore exchanged as $x \rightarrow x+2 \pi$:
\begin{eqnarray}
\lambda_\pm(\tau+2\pi,x)&=&\lambda_\pm(\tau,x)~,\nonumber\\
\lambda_\pm(\tau,x+2\pi)&=&\lambda_\mp(\tau,x)~.
\label{l2}
\end{eqnarray}
\par
Notice that in order to have an exchange of the eigenvalues in both the $x$ and
the $\tau $ direction it would be enough to replace at the r.h.s. of 
(\ref{esempio}) $x$ with $x+\tau$.
In conclusion every configuration of the field $F(\tau,x)$ belongs to a 
topological sector labelled by an ordered pair of commuting permutations. 
In the previous example $F(\tau,x)$  belongs to the $({\bf 1},Q)$
sector with $Q=(1,2)$.
In Appendix A we give an explicit construction of all pairs of commuting 
permutations.
We have already remarked that the transformation $g(\tau,x)$ that diagonalizes
$F$ is defined only up to an element of the Weyl group, 
namely that if  $g(\tau,x)$ diagonalizes $F$, then so does $R g(\tau,x)$ with 
$R\in S_N$. There is therefore a residual ambiguity in the gauge fixing which 
could be removed by fixing for instance the order of the eigenvalues at a 
specific point.
We can avoid doing that by simply dividing the functional integral by $N!$
to account for the multiplicity of the gauge equivalent copies. 
Notice that fields obeying  boundary conditions of the type (\ref{bcon}) 
and (\ref{abc}) are gauge equivalent to fields obeying  the same boundary 
conditions with $(P,Q)$ replaced by $(R PR^{-1}, R Q R^{-1})$.
Thus {\em gauge-inequivalent} topologically distinct sectors are in one-to-one 
correspondence  with pairs of {\em conjugacy classes} of commuting permutations. 
The functional integral over each sector gives a partition function $Z(t,P,Q)$,
that we shall evaluate shortly. The total partition function will be obtained by
summing over all sectors with suitable relative weights $c(P,Q)$:
\begin{equation}
Z(\Sigma_1,t)={1\over N!}\sum_{P,Q}c(P,Q)Z(t,P,Q)~,
\label{fullpart}
\end{equation}
where the factor $1/N!$ is inserted to account for the gauge ambiguity 
discussed above.
The problem of determining the weights $c(P,Q)$ will be discussed later.        
\par
The fundamental feature, shared by all the different sectors, is the exact
cancellation between the Faddeev-Popov determinant and the contribution of the
non-diagonal part of the gauge field $A_\mu$. As discussed in the previous
section this follows from the  supersymmetry (\ref{supersy}) which is unbroken
in case of zero curvature surfaces.
\par
As a result, we are left with the U$(1)^N$ invariant part of the action, which
now reads
\begin{eqnarray}
Z(P,Q,t)&=&\int\left(\prod_{i}[dA_\mu^{(i)}]
[d\lambda_{i}]\right)\nonumber\\
&&\exp\left\{- \int_0^{2\pi}d\tau dx\sum_{i}\left[\frac{t}{2} \lambda_{i}^2
-\ii\lambda_i \left( \partial_{0} A_{1}^{(i)}-\partial_{1} A_{0}^{(i)}\right)
\right]\right\}~.
\label{qeds}
\end{eqnarray}
This would be just $N$ copies of QED on a torus, except for the fact that the
$N$ copies are  mixed by the boundary conditions, which are of the type
described in Eq. (\ref{bcon})  for all the fields involved: 
\par
Of course in the trivial sector ($P=Q={\bf 1}$) the result is trivial and 
coincides with the $N$th power of the partition function of QED:
\begin{equation}
Z({\bf 1},{\bf 1},t)=\left(Z_{\rm QED}(t)\right)^N = \sum_{n_i} \exp \left( 
-2\pi^2 t
\sum_{i=1}^{N} n_i^2 \right)~,
\end{equation}
where the sum over the integers $n_i$ is unrestricted; in particular, coincident
values of different $n_i$'s are not excluded.
Let us proceed to study non-trivial sectors, by considering first a special
case, in which the permutation $P$ is given by $r_k$ 
cycles of length $k$, and $Q$ acts as a cyclic permutation of the
$r_k$ cycles of $P$. 
\begin{figure}
\begin {center}
 \null\hskip 1pt
 \epsfxsize 11cm 
 \epsffile{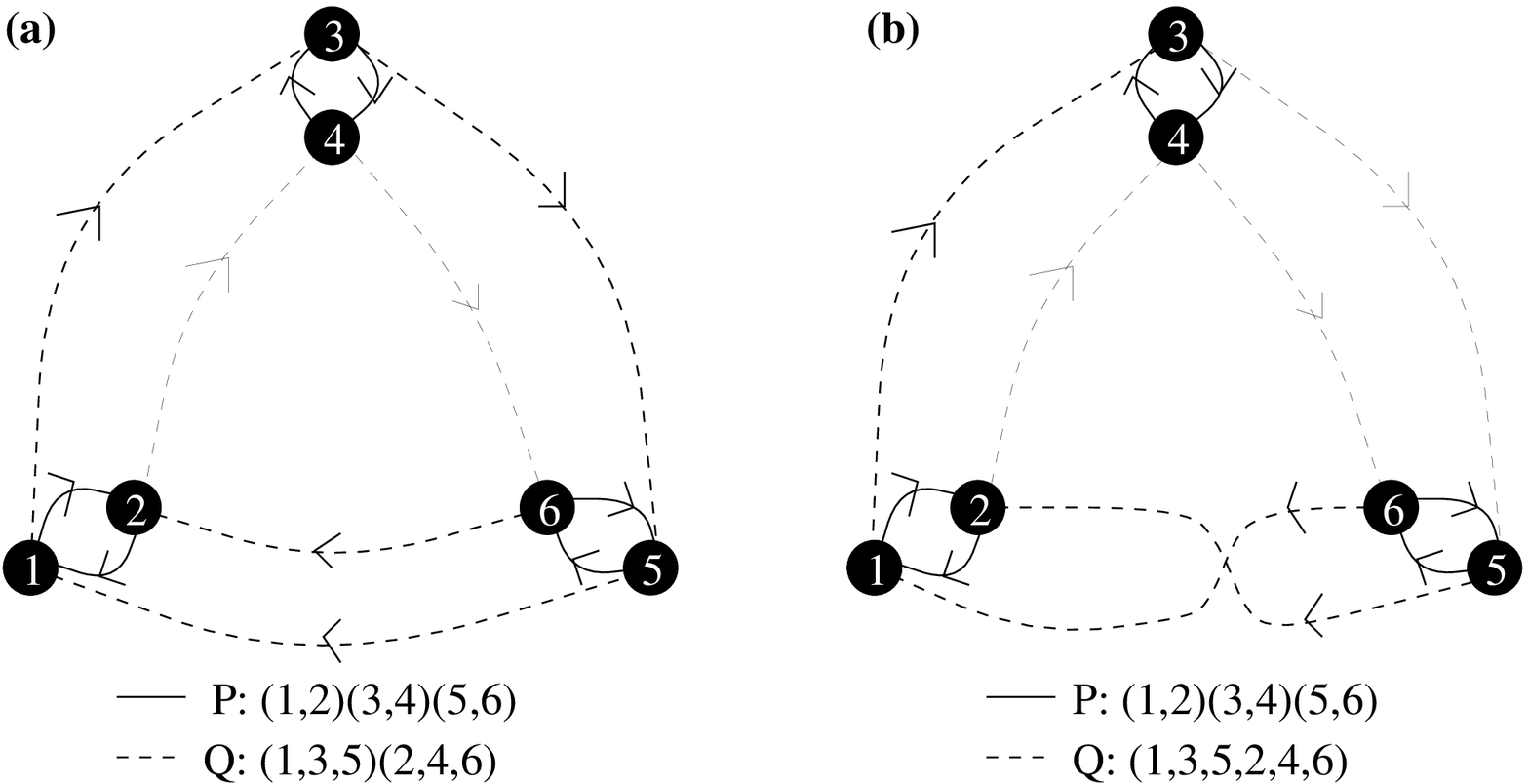}
\end{center}
\mycaptionl{A permutation $P$ and 
a commuting permutation $Q$ (dashed lines) consisting of   
2 cycles (a) or 1 cycle (b).} 
\end{figure}
An example, where $P$ consists of three cycles of length
two, is illustrated in Fig. 1, where the cycles of $P$ are represented by 
continuous lines joining the different points. Different choices for $Q$ are
given in Fig. 1(a,b) where the dotted lines represent the $Q$-cycles. The two
cases correspond to $Q$ consisting of 2 cycles of length 3 or 1 cycle of 
length 6. 
It is easy to convince oneself that in the situation described above $k r_k$
eigenvalues obeying the boundary conditions (\ref{bcon}) are equivalent to one  
eigenvalue satisfying the boundary conditions
\begin{eqnarray}
\lambda(\tau+2 k\pi,x)&=&\lambda(\tau,x)~,\\
\lambda(\tau,x+2 r_k \pi)&=&\lambda(\tau+2 S \pi,x)~,
\label{torogrande}
\end{eqnarray}
where $S$ is an integer shift, which in the notations of appendix A is given by
$S=\sum_\alpha s(k,\alpha)$. In the example of Fig. 1, this is illustrated by
Fig. 2, where the universal covering of the torus and the fundamental region
are represented (by dotted lines). The opposite sides of the fundamental region can be identified
only modulo a permutation of the eigenvalues. Fig.s 2(a,b) show the 
fundamental regions of a torus of area $k r_k$  corresponding to the boundary conditions
(\ref{torogrande}) in the cases of Fig.s 1(a,b).

In conclusion the partition function for a non-trivial sector with $P$ given by 
$r_k$ cycles of length $k$ and $Q$ acting as a permutation of such cycles 
coincides with the partition functions of QED
defined on a torus of area $kr_k$ times the original torus 
(the QED partition function on a torus does not depend on the modular
parameter of the torus but only on its area), namely
\begin{equation}
Z_{\rm QED}(kr_kt) = \sum_{n}  \exp \left( - 2\pi^2kr_k t\, n^2 \right)~.
\label{zqed}
\end{equation} 
\par
\begin{figure}
 \begin{center}
  \null\hskip 1pt
  \epsfxsize 11cm
  \epsffile{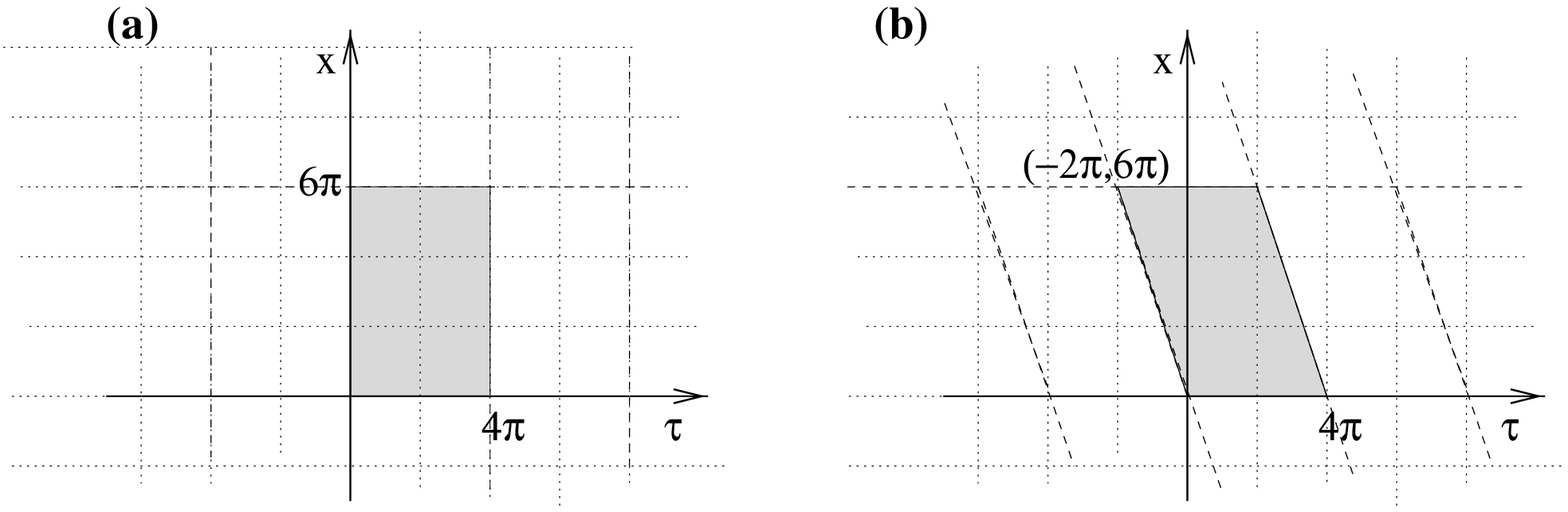}
 \end{center}
 \mycaptionl{(a,b): Tori on which the QED's corresponding 
 to the topological sectors
 $(P,Q)$ as in Fig. 1(a,b) are defined.}
\end{figure}
A  pair $(P,Q)$ of commuting permutations consists in general of several
blocks of connected cycles, like the one discussed above and pictured as an
example in Fig. 1(a,b). Correspondingly its partition function will consist of
the product of QED partition function defined on tori of area proportional to
the number of points in each block. For instance the sector corresponding to the
pair of permutations illustrated in Fig. 3 has a partition function given by:
\begin{equation}
\label{eqfig3}
Z_{\rm Fig.\,3}(t) = \sum_{n_1,n_2,n_3} \exp \left[ -2\pi^2 t (6 n_1^2 +5
n_2^2+n_3^2) \right]~.
\end{equation}
The general expression for the partition function of the $(P,Q)$ sector is
\begin{equation}
Z_{PQ}(t)=\prod_{k=1}^N\prod_{h=1}^{r_k}\left[Z_{\rm QED}(hkt)\right]^{s_h(k)}~,
\label{zpq}
\end{equation}
where the $r_k$ is the number of cycles of length $k$ in $P$. $Q$ acts on these
as a  permutation $\pi_k$ and the exponent $s_h(k)$ is the number of cycles of 
length $h$ in  $\pi_k$.
\par
The complete partition function is obtained as a sum over all different 
sectors 
with with weights $c(P,Q)$ according to Eq. (\ref{fullpart}). So the problem is
to determine to what extent the coefficients $c(P,Q)$ can be fixed from
consistency requirements. In principle the different  sectors correspond to
disconnected parts of the functional integral, and could be added with arbitrary
coefficients. It is shown in appendix B, that in the BRST invariant formulation
they correspond to gauge fixing functions which are non connected to each other,
so that BRST invariance does not tell us anything about their relative weights.
On the other hand we may require that the partition function is unchanged if we
perform a Dehn twist, or more generally a modular transformation, on the torus.
The generators of the modular group $\cal S$ and $\cal T$ act on a given sector 
$(P,Q)$
in the following way:
\begin{eqnarray}
{\cal S}:~~~~~~~  (P,Q)&\to&(Q,P)~,\\
{\cal T}:~~~~~~~  (P,Q)&\to&(PQ,Q)~.
\label{modtrans}
\end{eqnarray}
\begin{figure}
 \begin{center}
  \null\hskip 1pt
  \epsfxsize 8cm
  \epsffile{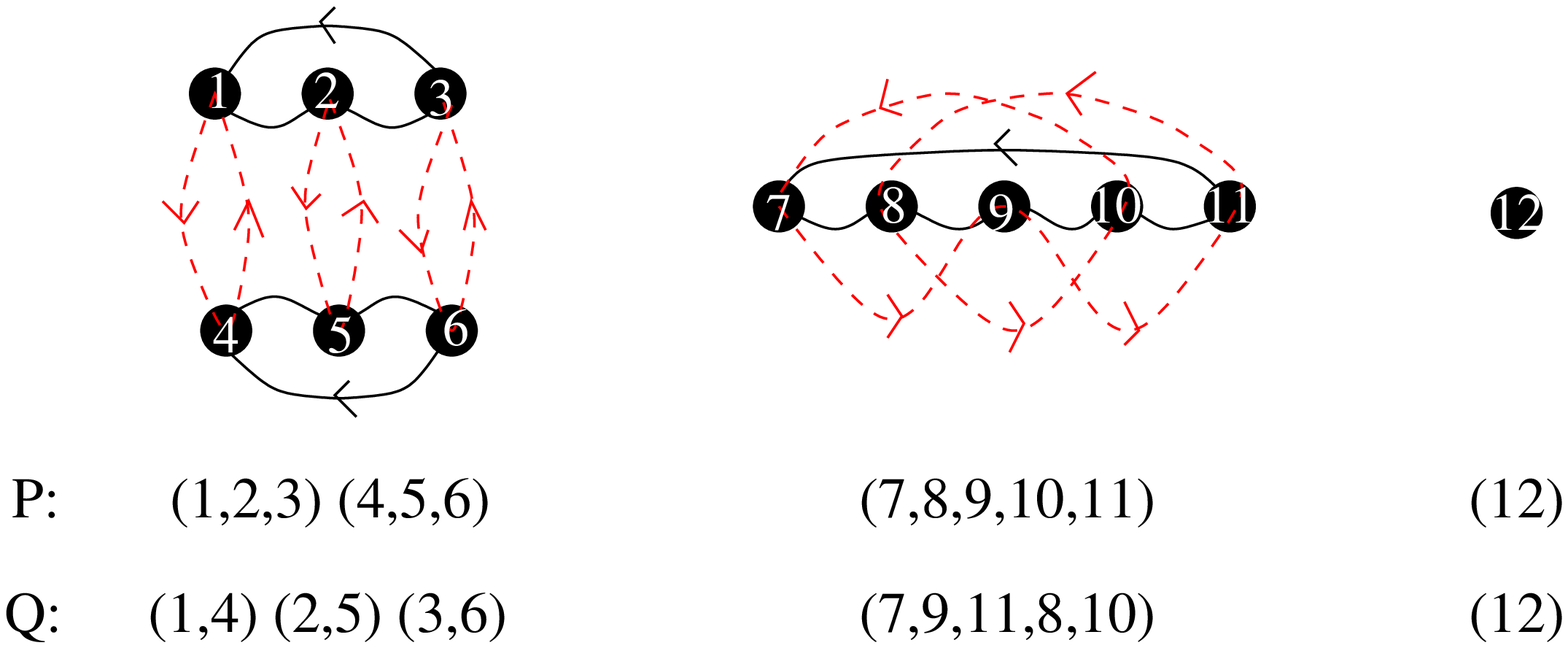}
 \end{center}
 \mycaptionl{A topological sector of the $SU(12)$ theory, defined by $P$
 (solid lines) and $Q$ (dashed lines), whose partition function is given in 
 Eq. (\ref{eqfig3}).}
\end{figure}
It is easy to check that the resulting pair of permutations still commute and
that the dimension of the connected blocks of cycles, which determines the
decomposition of the partition function in terms of QED partition functions,  
is left unchanged by modular transformations. For instance it is clear from 
Fig.s 2(a,b) that the blocks described in Fig.s 1(a,b) are obtained from each
other by a modular transformation of the torus. 
The invariance of (\ref{fullpart}) under the modular group implies then that 
$c(P,Q)=c(P',Q')$ if $(P,Q)$ and $(P',Q')$ are related by a modular
transformation.
One would also expect to recover the standard partition function found in the
literature by summing over a subset of sectors $(P,Q)$, namely the subset where
one permutation, say $P$, is the identity.This can be understood if one follows the standard derivation of the partition
function (see for instance Ref.~\cite{cdmp94}), which is obtained from the
kernel of the cylinder by identifying the holonomies at the borders and taking
the trace.
This automatically involves a sum over permutations $Q$, in fact the eigenvalues
of the holonomies are identified up to a permutation when the two edges are sewn
together.
There is no sign in this derivation of the sectors with a non trivial
permutation $P$ associated to the other cycle of the torus.
This problem will be analyzed in Section 4, where the theory on a cylinder that 
includes all $(P,Q)$ sectors is developed. We just anticipate here that non 
trivial permutations $P$ correspond to holonomies where the eigenvalues belonging 
to the same cycle of order $k$ of $P$ (namely on which $P$ acts as a cyclic 
permutation) are proportional to the $k$-th roots of unity as shown in Eq. 
(\ref{m4}). 
So within each sector the trace is an integral over a number of invariant angles 
equal to the number of cycles of $P$, and the standard group integration 
automatically projects over the trivial sector.
Even in the $P={\bf 1}$ sector an ambiguity is present when the trace over the
holonomies is taken. In the standard quantization this corresponds to an
integration over the group manifold and the wave functions of the states at the
edges of the cylinder are antisymmetric with respect to the exchange of the
eigenvalues.
Correspondingly a factor $(-1)^{|Q|}$ is obtained when the sum over the
permutations is taken, and its effect is to cancel all non  regular terms in the
partition function (see for instance Ref.~\cite{cdmp94} ).
It is also possible however to quantize over the algebra rather than over the
group. In this case the wave functions are symmetric and the result coincides
with the one given in Ref.~\cite{h94}.

\subsection{Modular invariant partition functions}
Let us consider the partition function 
\begin{equation}
Z_N(t)={1\over N!}\sum_{PQ}c(P,Q)Z_{PQ}(t)~,
\label{zn}
\end{equation}
where the  coefficients $c(P,Q)$ satisfy the requirements of modular invariance
\begin{eqnarray}
c(P,Q)&=&c(Q,P)~, \nonumber \\
c(P,Q)&=&c(PQ,Q)~.
\label{modin}
\end{eqnarray}
It is more convenient to work directly on the grand canonical partition function 
defined by
\begin{equation}
Z(t,q)=\sum_N
Z_N(t) q^N~.
\label{gc}
\end{equation}
If one imposes on $c(P,Q)$ only the constraint (\ref{modin}) of modular
invariance, the partition function (\ref{zn}) has as many free parameters
as the number of commuting permutation not related by a modular transformation.
In order to further restrict the possible choices we shall consider the case 
where $c(P,Q)=\pm 1$ for all pairs $(P,Q)$. As we shall see, this leads to 
partition functions that in the sub-sector where $Q=1$ coincide with the 
standard partition function on the torus or with the one obtained 
by quantizing on the algebra rather than the group \cite{h94}.
The simplest case is when $c(P,Q)=1$. In order to calculate $Z(t,q)$ in 
this case, let us review some combinatorial formulas. The number of
permutations $P$ with a given structure in cycles, namely with $r_k$
cycles of order $k$, is given by $N!/\prod_k (r_k! k^{r_k})$ and 
the number of permutations $Q$ commuting with $P$ are $\prod_k r_k! k^{r_k}$.
As shown in Appendix A, $Q$ acts, for each $k$, as a permutation of the $r_k$
cycles of order $k$ in $P$. Let $s_h(k)$ be the number of cycles of order $h$
in such permutation. The set of numbers $s_h(k)$ characterizes completely the 
decomposition  into connected blocks of the pair $(P,Q)$. The number of
pairs $(P,Q)$ corresponding to a given choice of $s_h(k)$ can be easily 
calculated
and is given by $N!/\prod_{h,k} [s_h(k)! h^{s_h(k)}]$. In conclusion
the grand canonical partition function (\ref{gc}) can be written as
\begin{equation}
Z_{\rm b}(t,q)= \prod_{h,k} \sum_{s_h(k)} \frac{ q^{h k s_h(k)} 
\left( \sum_n \esp{-2\pi^2 \,hkt n^2} \right)^{s_h(k)} }{s_h(k)! h^{s_h(k)}}~.
\label{gc2}
\end{equation}
The sums over $s_h(k)$ and $h$ can be done explicitly, leading to the result
\begin{equation}
Z_{\rm b}(t,q)= \prod_{n=-\infty}^{\infty} 
\prod_{k=1}^{\infty} \frac{1}{1-q^k e^{-2\pi^2 kt\,n^2}}~,
\label{gc3}
\end{equation}
which can be interpreted as the grand canonical partition function of a
collection of free bosons: 
\begin{equation}
Z_{\rm b}(t,q)  = \Tr\, q^{N_c} \esp{-t E}~,
\end{equation}
where the trace is defined on an Hilbert space generated by harmonic 
oscillators\footnote{The harmonic oscillators are normalized
by $[a_k(n),a_h(m)^{\dagger}] = \delta_{hk} \delta_{nm}$.}
$a_k(n)$ and $a_k(n)^{\dagger}$ 
and the operators $N_c$ and $E$ are defined by
\begin{eqnarray}
N_c&=& \sum_{k,n} k a_k(n)^{\dagger} a_k(n)~,\nonumber \\
E&=&\sum_{k,n} k n^2 a_k(n)^{\dagger} a_k(n)~.
\label{ener}
\end{eqnarray}

$Z_{\rm b}(t,q)$ can be rewritten as an infinite product of Dedekind functions,
with modular parameters $\tau_n$ which are functions of $n$:
\begin{equation}
Z_{\rm b}(t,q)= \esp{\alpha_0}\prod_{n=-\infty}^{\infty} [\eta(\tau_n)]^{-1}
\label{gcded}
\end{equation}
with 
\eq
\eta(\tau)= \esp{\frac{\ii\pi \tau}{12}}
\prod_{k=1}^{\infty} (1-\esp{2\pi \ii\tau k })
\en
and
\eq
\tau_n=\ii(\mu+\pi t n^2)~,
\en
where $\mu$ is the chemical potential, defined by $q\equiv \esp{-2\pi\mu}$ and 
\eq
\alpha_0=-\frac{\pi}{12}\sum_{n=-\infty}^{\infty}(\mu+\pi t n^2)
\en
This sum is divergent and must be regularized. Remarkably the zeta function 
regularization 
gives just $\alpha_0=0$, and our partition function becomes exactly an infinite
product of Dedekind functions. 

If we restrict the permutation $P$ in the $x$ direction to be the identity,
namely we restrict the product in (\ref{gc3}) to $k=1$, the expansion of the 
r.h.s.
of (\ref{gc3}) in powers of $q$ reproduces the partition
function on a torus obtained by Hetrick in \cite{h94} by quantizing YM2 on 
the algebra rather than on the group. 
\par
Let us consider now the partition functions where $c(P,Q) =\pm 1$.
These are obtained by inserting in the sum in (\ref{gc2}) a sign $(-1)^{f}$,
where $f$ is an integer and is a modular invariant function of $h$, $k$ and
$s_h(k)$. There are two such modular invariant quantities one can construct:
$\sum_{h,k} h k s_h(k)=N$  and $\sum_{h,k} s_h(k)$. 
The latter is the number of the connected blocks of cycles in 
the given sector. Both quantities are preserved by modular transformations.
The introduction in (\ref{gc2}) of a factor 
$(-1)^{\sum_{h,k} h k s_h(k)} =(-1)^N$ just changes the overall sign of the 
partition functions with odd values of $N$;  in the grand canonical partition
function it is equivalent to the substitution $q \rightarrow -q$.
The insertion of a factor $(-1)^{\sum_{h,k} s_h(k)}$ is more interesting as it 
turns the bosonic partition function (\ref{gc3}) into a fermionic one:
\begin{equation}
Z_{\rm f}(t,q)=\prod_{n=-\infty}^{\infty} 
\prod_{k=1}^{\infty} \left(1-q^k \esp{-2\pi^2 kt\,n^2} \right)= 
\prod_{n=-\infty}^{\infty} [\eta(\tau_n)]~.
\label{gc4}
\end{equation}
$Z_{\rm f}(t,q)$ can be written as a trace on a Hilbert space generated by 
fermionic (anti-commuting) oscillators $b_k(n)$ and $b_k(n)^{\dagger}$
\begin{equation}
Z_{\rm f} = \Tr\, (-1)^F q^{N_c} \esp{-tE}~,
\end{equation}
where the operators $F$ (fermionic number),$N_c$ 
and $E$ are given by
\begin{eqnarray}
N_c &=& \sum_{k,n} k b_k(n)^{\dagger} b_k(n)~,\nonumber \\
E&=&\sum_{k,n} k n^2 b_k(n)^{\dagger} b_k(n)~, \\
F&=&\sum_{k,n}  b_k(n)^{\dagger} b_k(n)~.\nonumber 
\label{ener2}
\end{eqnarray}
The restriction to $P=1$ leads in this case to the standard partition
function for YM2 on a torus. This case was already discussed in 
\cite{cdmp94}, and the equivalence of the standard approach with the 
present formulation restricted to $k=1$ can be seen by comparing
(\ref{gc4}) with Eq. (41) of Ref. \cite{cdmp94}\footnote{The slight discrepancy
between the two expressions in the case of even $N$ is due to a different 
coupling of the U$(1)$ factor within the U$(N)$ group. In fact the discrepancy 
disappears in the case of SU$(N)$ where the quantized momentum $n$ is shifted by 
the ``center of mass momentum" $\beta$  and an integration over $\beta$ is 
included in the definition of the trace. This was discussed in Ref.~\cite{cdmp94} 
and it can be shown that the same prescription applies in the present 
generalization.}.
\par
A particularly interesting limit, in both fermionic and bosonic partition 
function, is $t \rightarrow \infty$. This is the limit where the matrix string 
theory of Ref.~\cite{dvv} has an infrared fixed point described by a conformal 
field theory.
In this limit the free string is recovered as the string coupling $g_s$ is 
essentially $1/t$. For $t \rightarrow \infty$ only the $n=0$ excitations survive 
in Eq.s (\ref{gc3}) and (\ref{gc4}). In the case of $Z_{\rm b}(q,t)$  we recover 
(apart from an exponential prefactor) the partition function of the conformal 
field theory of a single boson living on a rectangle with Dirichlet boundary 
conditions and a ratio $\mu$ between the two sides. However this rectangle has 
nothing to do with the torus on which the original YM theory is defined.

Another interesting limit is $t \rightarrow 0$. In this limit Yang-Mills theory
becomes a BF theory and  $Z_{\rm b}(q,t)$ becomes formally an infinite product of
partition functions identical to the one discussed in the $t \rightarrow \infty$
limit. This clearly shows that the limit is singular. The singularity can be 
handled by using in (\ref{gc2}) the Poisson summation formula and writing
\begin{equation}
 \sum_n \esp{-2\pi^2 \,hkt n^2}=\frac{1}{\sqrt{2 \pi hkt}} + O(\esp{-\frac{\rm 
const}{t}})
~~~~~~~~~~~~~~~~~~~~~~(t \rightarrow 0)~,
\label{tzero}
\end{equation}
which implies
\begin{equation}
Z_{\rm b,f}(t,q) = \esp{\pm \frac{1}{\sqrt{2 \pi t}} \sum_{h,k} 
\frac{q^{hk}}{h^{3/2}k^{1/2}}
+  O(\esp{-\frac{\rm const}{t}})}~,
\label{tzero2}
\end{equation}
where the $+$ and $-$ sign at the exponent refer to $Z_{\rm b}$ and $Z_{\rm f}$ 
respectively.
It is apparent from (\ref{tzero}) and (\ref{tzero2}) that a $1/\sqrt{t}$ 
singularity is associated to each connected block in the $(P,Q)$ into cycles. 
This means that at fixed $N$ the leading most singular term is of order 
$t^{-\frac{N}{2}}$ and comes from the 
$(P={\bf 1},Q={\bf 1})$ sector, namely from cycles of order 1.
On the contrary in the $t \rightarrow \infty$ limit the mean value at fixed $N$ 
of the length of a cycle can be estimated ~\cite{peri} and found to be larger 
that $O(\sqrt{N} \log \sqrt{N})$.
This might be a signal that in the large $N$ limit  at  some critical value 
$t_c$ a phase transition occurs from a short cycle to a long cycle regime.

>From the physical point of view the situation can be described as follows:
we have two types of degrees of freedom,  the momentum excitations labelled by 
$n$ and the string degrees of freedom labelled by the
length $k$. Correspondingly we have two free
parameters: the YM coupling $t$ and the chemical potential $\mu$ that set the 
mass scale for the corresponding excitations\footnote{Note that while the 
dependence of the states' energy from $k$ is fixed, the dependence from $n$ 
reflects the form of the $\tr\, F^2$ term in the original action
(\ref{partriemann}). Replacing $\tr\, F^2$  with the trace of an arbitrary
potential $V(F)$ would amount to substitute $n^2$ with
$V(n)$ in the partition functions.}. From the point of view of a Matrix string 
theory interpretation ~\cite{dvv}, $t \simeq 
1/g_s$  is the inverse of the string coupling constant.
In the strong coupling limit of the string ($t \rightarrow 0$) the string breaks 
up into a gas of partonic constituents (the $k=1$ states) while in the weak 
coupling regime the tension effects prevail and long strings are energetically 
favoured. 
Notice that the chemical potential $\mu$ was not present in the original YM 
theory, but it was introduced because our results have a natural interpretation 
in terms of the grand canonical partition function. Its introduction from the 
very beginning would amount to writing the U$(N)$gauge action as
\begin{equation}
S_N(t,\mu) = \tr_{N} \int dx\, d\tau \left(\frac{t}{2} F_{(N)}^2 - \ii 
f(A_{(N)})F_{(N)}
\right) + 2 \pi \mu N~,
\label{acpc}
\end{equation}
where the labels $N$ are to denote the dimension of the matrices.
The partition function is then defined by
\begin{equation}
Z(t,\mu) = \sum_N \int [dA_{(N)}][dF_{(N)}] \esp{-S_N(t,\mu)}~.
\label{pfcpc}
\end{equation}
This establishes a close analogy with the IKKT matrix string theory for type IIB
strings ~\cite{IKKT} where a similar sum over $N$ is involved. 

\section{Path integral on the cylinder}
\label{sec4}
Let us consider  the path-integral (\ref{partriemann}) with $\Sigma_g$
a cylinder, that we can represent as a square of area $4\pi^2$, with periodic
identification in the space-like $x$ direction. In this section we shall 
perform the calculation of the functional integral and derive the kernel on the
cylinder as a function of the degrees of freedom at the edges.
Finally, by sewing the two edges of the cylinder together we shall reproduce
the partition functions on a torus obtained in the previous section.

As discussed in Section \ref{sec3}, we fix the unitary gauge by 
performing the gauge transformation $g(\tau,x)$ that diagonalizes $F$. 
The continuity of $g(\tau,x)$ leads one to consider the generalized 
boundary conditions
\begin{equation}
 \label{cyl1}
 g(\tau,x+2\pi) = Q g(\tau,x)~,
\end{equation} 
where $Q$ is a permutation. Thus in the unitary gauge the gauge fields 
$A^{\rm (u)}_\mu$ and the eigenvalues $\lambda_i$ of $F$ satisfy 
\begin{equation}
 \label{cyl2}
 A^{\rm (u)}_\mu(\tau,x+2\pi) = Q^{-1} A^{\rm (u)}_\mu(\tau,x) Q~,
\end{equation}
\begin{equation}
 \label{cyl3}
 \lambda_i(\tau,x+2\pi) = \lambda_{Q(i)}(\tau,x)~.
\end{equation}
It is convenient to write the last condition using for the index $i$  the 
multi-index notation introduced in Appendix A: $i\to (k,\alpha,n)$ where the
set of three indices label the $n$-th element of the $\alpha$-th cycle of 
length $k$ in Q. The range of the indices is then  $\alpha = 1,\ldots r_k$ 
with $\sum k r_k = N$, and $n=1,\ldots k$.
In this notation Eq. (\ref{cyl3}) reads 
\begin{equation}
 \label{cyl3bis}
  \lambda_{k,\alpha,n}(\tau,x+2\pi) = \lambda_{k,\alpha,n+1}(\tau,x)~.
\end{equation} 
where here and in the following the index $n$ is understood  ${\rm mod}\, k$. 

In order to understand the effect of the non trivial boundary conditions
(\ref{cyl3bis}), let us first study the topologically non-trivial Wilson loop
\begin{equation}
W(\tau) \equiv P\,\exp\{-\ii \int_0^{2\pi} dx A_1(\tau,x)\} 
\in {\rm U}(N)
\label{wilson}
\end{equation}
and denote by $W^{\rm (u)}(\tau)$ its expression in 
the unitary gauge. $W(\tau)$ and $W^{\rm (u)}(\tau)$ are related
by the gauge transformation $g(\tau,x)$ taken at the end points $x=0$ and 
$x= 2 \pi$:
\begin{equation}
 \label{cyl3bb}
 W(\tau) = g^{-1}(\tau,0) W^{\rm (u)}(\tau) g(\tau,2\pi) =  
 g^{-1}(\tau,0) W^{\rm (u)}(\tau) Q g(\tau,0)~,
\end{equation}
According to Eq. (\ref{cyl3bb}), the eigenvalues 
$\esp{\ii \theta_i(\tau)}$ of $W(\tau)$ 
coincide with the eigenvalues of $W^{\rm (u)}(\tau) Q $. 
In the unitary gauge, on the other hand, the non diagonal matrix 
elements of $A^{(u)}_1(\tau,x)$ are forced to vanish as a result of the 
functional integral over $A^{\rm (u)}_0(\tau,x)$ with the
action (\ref{offdiag}), and $W^{\rm (u)}(\tau) $ is therefore diagonal.
It is easy to see that with $W^{\rm (u)}(\tau) $  diagonal the matrix 
$W^{\rm (u)}(\tau) Q$
has  in the multi-index notation  the form
\begin{equation}
(W^{\rm (u)}(\tau) Q)_{k,\alpha,n;k',\alpha',n'} = 
\delta_{k,k'}\delta_{\alpha,\alpha'}\delta_{n,n'-1}
\esp{\ii \phi_{k,\alpha,n}}~,
\label{wilson2}
\end{equation}
where $\phi_{k,\alpha,n}$ are the invariant angles of  $W^{\rm (u)}(\tau)$.
The eigenvalues of the matrix at the r.h.s. of (\ref{wilson2}) can be
easily calculated to be
\begin{equation}
\theta_{k,\alpha,n} = \theta_{k,\alpha}+\frac{2\pi\ii\,n}{k}~,
\label{holo}
\end{equation}
with 
\begin{equation}
\label{cyl4bbb}
\theta_{k,\alpha}(\tau)  = {1\over k} \sum_{n=1}^k {\phi}_{k,\alpha,n}(\tau) =
-{1\over k} \sum_{n=1}^k \int_0^{2\pi} dx A_1^{k,\alpha,n}(\tau,x)~,
\end{equation}
where $ A_\mu^{k,\alpha,n}(\tau,x)$ are the diagonal elements of the 
gauge fields in the unitary gauge. 
In conclusion, the eigenvalues of the original Wilson loop $W(\tau,x)$ are 
not independent; rather, for each cycle of length $k$ of $Q$ they are distributed 
as the $k$-th roots of unity shifted by a common value $\theta_{k,\alpha}$
which is defined modulo $2\pi/k$ instead of modulo  $2\pi$. 

Let us go back to the functional integral (\ref{partriemann}), and observe that 
due to the cancellation of the Vandermonde determinants as a result of the 
fermionic symmetry (\ref{supersy}),  we are left with a collection of QED-type 
actions as in Eq. (\ref{qeds}).
The fields in (\ref{qeds}) whose U$(N)$ index belong to the same cycle
in the cycle decomposition of $Q$, are related to each other by the boundary 
conditions (\ref{cyl3bis})  and they can be reduced to one field 
$\tilde\lambda_{k,\alpha}(\tau,x) $ with $x$ ranging in the
interval $(0,2\pi k)$ instead of $(0,2\pi )$:
\begin{equation}
 \label{cyl5}
 \tilde\lambda_{k,\alpha}(\tau,x) = \left\{
 \begin{array}{ll} 
 \lambda_{k,\alpha,1}(\tau,x)~, &  0\leq x < 2\pi~, \cr 
 \lambda_{k,\alpha,2}(\tau,x-2\pi)~, &  2\pi\leq x < 4\pi~, \cr 
 \ldots & \ldots \cr
 \lambda_{k,\alpha,k}(\tau,x-2(k-1)\pi)~, & 2(k-1)\pi\leq x < 2k\pi~.
 \end{array}\right.
\end{equation}
Similarly a U$(1)$ gauge field  $\tilde{A}^{k,\alpha}_\mu(\tau,x)$ with period
$2\pi k$ in $x$ can be defined from $A^{k,\alpha,n}_\mu(\tau,x)$.
In conclusion, the functional integral on the cylinder for the sector 
corresponding to
a permutation $Q$ decomposes into a product of functional integrals, one for each
cycle of $Q$, with the action being the one of a QED defined on a cylinder of 
length $2\pi k$ in the compactified direction:
\begin{eqnarray}
 \label{cyl6}
 &&Z^{\rm cyl}(t) =
 \prod_k  \int\prod_{\alpha=1}^{r_k} [d\tilde{A}_\mu^{k,\alpha}]
 [d\tilde\lambda_{k,\alpha}]\nonumber\\
 &&\exp\left\{- \sum_{\alpha=1}^{r_k}\int_0^{2\pi}d\tau \int_0^{2\pi k} dx
 \left[\frac{t}{2} \tilde\lambda_{k,\alpha}^2
 -\ii\tilde\lambda_{k,\alpha} \left( \partial_{0} \tilde{A}_{1}^{k,\alpha}-
 \partial_{1} \tilde{A}_{0}^{k,\alpha}\right)
 \right]\right\}~,
\end{eqnarray}
where $k$ is the length of the cycle.
As discussed in Section \ref{sec3}, after Eq. (\ref{l2}), 
the sum over the sectors involves a further gauge fixing related to the
fact that the diagonal gauge is defined up to an arbitrary permutation of the
eigenvalues: if $g(\tau,x)$ is a gauge transformation that
diagonalizes $F$, so is $R g(\tau,x)$ with $R$ an arbitrary permutation. It
satisfies
\begin{equation}
\label{corr1}
R g(\tau,x+2\pi) = RQR^{-1}~Rg(\tau,x)~, 
\end{equation}
which show that sectors characterised by permutations $Q$ and $Q'=RQR^{-1}$,
belonging to the same conjugacy class, are gauge equivalent. This implies that
the sum over all sectors involves a sum over the  conjugacy classes rather that
a sum over the permutations $Q$.
Even so there is still a residual gauge transformation given by the permutations
$R$ that commute with $Q$  (i.e. $R\in C(Q)$, $C(Q)$ being
called the centralizer of $Q$. In the following we shall denote with $P$
a generic permutation belonging to $C(Q)$). 
As described in Appendix A, such a permutation P acts on the multi-index 
$(k,\alpha,n)$ by
\begin{equation}
 \label{ncyl2}
(k,\alpha,n)\stackrel{P}{\longrightarrow} (k,\pi_k(\alpha),n+ s(k,\alpha))~,
\end{equation}
where $\pi_k\in S_{r_k}$ is a permutations of $r_k$ elements and $s(k,\alpha)$
is an integer ${\rm mod}\, k$. It follows from this equation and
the definition (\ref{cyl5}) of $\tilde\lambda_{k,\alpha}(\tau,x)$, that the
gauge transformation $P$ acts on $\tilde\lambda_{k,\alpha}(\tau,x) $ and
$ \tilde{A}^{k,\alpha}_\mu(\tau,x)$  in the following way:
\begin{eqnarray}
 \label{cyl7}
 \tilde\lambda_{k,\alpha}(\tau,x) & \stackrel{P}{\longrightarrow} &
 \tilde\lambda_{k,\pi_k(\alpha)}(\tau,x-2\pi s(k,\alpha))~,
 \nonumber\\
 \tilde{A}^{k,\alpha}_\mu(\tau,x) & \stackrel{P}{\longrightarrow} &
 \tilde{A}^{k,\pi_k(\alpha)}_\mu(\tau,x-2\pi s(k,\alpha))~.
\end{eqnarray} 

Also the eigenvalues of the Wilson loop given in Eq.(\ref{holo}) and 
(\ref{cyl4bbb}) can be expressed in terms of the redefined fields 
$\tilde{A}^{k,\alpha}$:
\begin{equation}
 \label{cyl7b}
k~\theta_{k,\alpha,n}(\tau)= k~\theta_{k,\alpha}(\tau) +2\pi n =
-\int_0^{2\pi k} dx \tilde{A}_1^{k,\alpha}(\tau,x)+~2\pi n~.
\end{equation} 

Let us proceed now to calculate the QED functional integrals in (\ref{cyl6})
by using a standard procedure (see \cite{cdmp94}).
We expand the fields appearing in (\ref{cyl6})
in their Fourier components in the compact $x$ 
direction: $\tilde{A}^{k,\alpha}_\mu(\tau,x) = $ 
$\sum_m \tilde{A}^{k,\alpha}_{\mu,m}(\tau) \exp(\ii m x/k)$, 
and similarly for $\tilde\lambda_{k,\alpha}(\tau,x)$. The U$(1)$ gauge
is fixed by choosing a Coulomb gauge $\partial_1 \tilde{A}^{k,\alpha}_1 = 0$, 
so that the only non-vanishing Fourier component of $\tilde{A}^{k,\alpha}_1$ is 
the zero mode which coincides, according to Eq. (\ref{cyl7b}), with 
$-\theta_{k,\alpha}(\tau)/(2\pi)$. 
The functional integration over $\tilde{A}^{k,\alpha}_0$ 
and the Gaussian integration over
the zero-mode of $\tilde\lambda^{k,\alpha}$ are straightforward and we remain
with
\begin{equation}
 \label{cyl10}
 Z^{\rm cyl}(Q,t)=\prod_k \int\prod_{\alpha=1}^{r_k}[d\theta_{k,\alpha}]
 \exp\left\{-{k\over 4\pi t} \sum_{\alpha=1}^{r_k}\int_0^{2\pi}d\tau
 (\partial_\tau \theta_{k,\alpha})^2\right\}~.
\end{equation} 
For each length $k$ of the cycle $Z^{\rm cyl}(Q,t)$ describes 
the quantum mechanics of $r_k$ free particles of mass $\mu=k/(2\pi t)$, 
that move on a circle of radius $2\pi/k$.
In fact, according to Eq. (\ref{holo}) and following discussion the coordinates
$\theta_{k,\alpha}(\tau)$ are defined modulo $2\pi/k$. 
Given the boundary conditions at $\tau=0$ and $\tau=2\pi$, 
namely $\theta_{k,\alpha}(0)$ and $\theta_{k,\alpha}(2\pi)$, 
the transition amplitude from the initial to the 
final configuration can be computed from (\ref{cyl10}) 
by using the methods described in
\cite{cdmp94}. We have:
\begin{eqnarray}
 \label{cyl14}
 &&{\cal K}_Q(\theta_{k,\alpha}(0),\theta_{k,\alpha}(2\pi) )=
 {1\over (2\pi)^N} \prod_k \left({k\over 2\pi}\right)^{r_k\over 2}\nonumber\\ 
 && \sum_{l_{k,\alpha}} 
 \exp\left\{-{k\over 8\pi^2 t} \sum_{\alpha=1}^{r_k} \left(
 \theta_{k,\alpha}(2\pi)-\theta_{k,\alpha}(0) - \frac{2\pi l_{k,\alpha}}{k} 
 \right)^2\right\}~,
\end{eqnarray}
where the sum over the winding numbers 
different
$k^{r_k/2}$
$l_{k,\alpha}$ ensures the periodicity in the configuration space of 
the $\theta_{k,\alpha}$'s. 
The sums over $l_{k,\alpha}$ can be performed by using the well known modular
transformation for the function $\theta_3$ 
(see for instance Eq. (28) in \cite{cdmp94}):
\begin{eqnarray}
\label{mom}
&&{\cal K}_Q(\theta_{k,\alpha}(0),\theta_{k,\alpha}(2\pi) )= 
\prod_k \left({k\over 2\pi}\right)^{r_k}\nonumber \\
&&
\sum_{n_{k,\alpha}} 
\exp\left\{-\sum_{\alpha=1}^{r_k} -2\pi^2 kt\,n_{k,\alpha}^2 -
\ii \sum_{\alpha=1}^{r_k} k n_{k,\alpha} 
\left( \theta_{k,\alpha}(2\pi)-\theta_{k,\alpha}(0)\right) \right\}~,
\end{eqnarray}
where the integers $n_{k,\alpha}$ can be interpreted 
as discretized momenta of the particles moving in the compactified 
configuration space. This is rather straightforward 
in the Hamiltonian formalism. In fact from (\ref{cyl10}) we find the 
Hamiltonian
\begin{equation}
H_Q = - \sum_k  {\pi t\over k} \sum_{\alpha=1}^{r_k}
(\partial / \partial\theta^{k,\alpha})^2
\label{hamil}
\end{equation}
which, due to the periodicity on $\theta^{k,\alpha}$, has discrete energy
levels:
\begin{equation}
E (n_{k,\alpha}) = \sum_k \pi k t  \sum_{\alpha=1}^{r_k} n_{k,\alpha}^2~,
\label{enlev}
\end{equation}
in agreement with Eq. (\ref{mom}).
Finally we observe that the residual gauge symmetry under  permutations $P$
that commute with $Q$ given in Eq. (\ref{cyl7}) reduces
to a permutation symmetry among the coordinates of the $r_k$ indistinguishable 
particles:
\begin{equation}
 \label{cyl12}
 \theta_{k,\alpha}(\tau) \stackrel{P}{\longrightarrow} 
 \theta_{k,\pi_k(\alpha)}(\tau)~.
\end{equation} 
The configuration space is then an orbifold with respect to the permutation 
group $S_{r_k}$:
\begin{equation}
 \label{cyl13}
 (S^1)^{r_k}/S_{r_k}~.
\end{equation}
It is consistent to think of these particles both as bosons or as fermions,
and we shall consider the two cases in the next subsection where we shall sew 
the cylinder to get the path-integral on the torus.

\subsection{Sewing the cylinder}
The partition function on the torus, studied in Section 3, can be 
reproduced from the results of the previous subsection by sewing the
two ends of the cylinder, that is by imposing periodicity  also 
in the $\tau$ direction. This has to be done keeping in due account 
the residual gauge invariance generated by the permutations that 
commute with $Q$.
In the Hamiltonian language the partition function on the torus
is given as a finite temperature trace:
\begin{equation}
\label{extra2}
Z(t) = \sum_{\{Q\}} {\rm Tr} (\esp{-\beta H_{Q}}\, {\cal P}_{Q})~.
\end{equation}
where $\beta$ is the inverse temperature and it is given in our
case by $\beta=\Delta \tau = 2\pi$. We shall consider both bosonic and fermionic
partition functions. In the former case  ${\cal P}_{Q}$ is just a
projection operator onto the states that are invariant under  
the permutations $P$ that commute with $Q$. This corresponds to 
projecting over states whose wave functions are completely symmetric under
(\ref{cyl12}). In the fermionic case two modifications are required:
the wave functions are chosen to be antisymmetric and a factor $(-1)^F$
counting the number of fermions is included in the trace. In our case $F$
is the number of antisymmetrized wave functions and so $(-1)^F = (-1)^{\sum 
r_k}$.
Calculating the trace at the r.h.s. of (\ref{extra2}) is the same as
identifying in Eq. (\ref{mom}) $\theta_{k,\alpha}(0)$ and 
$\theta_{k,\alpha}(2 \pi)$ up to an arbitrary permutation $\pi_k(\alpha)$  
of the index $\alpha$ coming from the (anti)symmetrization of the wave functions
and then integrating over the $\theta_{k,\alpha}$'s.
In fact Eq. (\ref{extra2}) can be rewritten in terms of the normalized wave 
functions
\begin{equation}
\braket{\theta_{k,\alpha}}{n_{k,\alpha}} =  \prod_k \left( \frac{k}{2 \pi} 
\right)^{r_k \over 2}
\esp{-\ii \sum_{\alpha=1}^{r_k} k n_{k,\alpha} \theta_{k,\alpha}}
\label{norwf}
\end{equation}
 as
\begin{equation}
\label{extra3}
 Z(t) =\sum_{\{Q\}} \prod_{k,\alpha} \int_0^{\frac{2 \pi}{k}} d\theta_{k,\alpha} 
\sum_{n_{k,\alpha}}
\braket{\theta_{k,\alpha}}{n_{k,\alpha}}
\bra{n_{k,\alpha}}\esp{-2\pi H_Q}\ket{n_{k,\alpha}}
\bra{n_{k,\alpha}}{\cal P}_{Q}\ket{\theta_{k,\alpha}}~.
\label{extra4}
\end{equation}
The integrand at the r.h.s. of (\ref{extra4}) is exactly the r.h.s. of 
(\ref{mom}) with the ends of the cylinder identified up to the effect of the 
projection operator ${\cal P}_{Q}$ which is in the bosonic case to symmetrize 
the wave function:
\begin{equation}
\bra{n_{k,\alpha}}{\cal P}_{Q}\ket{\theta_{k,\alpha}}=
\braket{n_{k,\alpha}}{\theta_{k,\alpha}}_{s}
= \prod_k\frac{k^{r_k/2}}{(2 \pi)^{r_k/2} r_k!} 
\sum_{\pi_k \in S_{r_k}} \esp{- 
\ii \sum_{k,\alpha} k n_{k,\alpha} \theta_{k,\pi_k(\alpha)} }~.
\label{extra5}
\end{equation}
In order to obtain the fermionic partition function the wave function has to be 
antisymmetrized,
namely\footnote{We can replace the antisymmetrized wave function 
$\braket{n_{k,\alpha}}{\theta_{k,\alpha}}_{a} $
with the ratio 
$\braket{n_{k,\alpha}}{\theta_{k,\alpha}}_{a}/[\prod_k J(k \theta_k)]$,
where $J(k \theta_k)$ is the Vandermonde determinant for unitary matrices:
$J(k \theta_k)=\prod_{\alpha<\beta} 
2 \sin [(k\theta_{k,\alpha} -k\theta_{k,\beta})/2]$. 
The integration over $\theta_{k,\alpha}$ should then be 
done with an integration volume
$J^2(k \theta_k) \prod_{\alpha} k d\theta_{k,\alpha}$ for each $k$, namely with 
the Haar measure of SU$(r_k)$
with invariant angles $k \theta_{k,\alpha}$. This corresponds, in the trivial 
sector $Q=1$, to fixing the holonomies
at the edge of the cylinder and doing a group invariant integration when the two 
edges are sewn together
(see for instance \cite{cdmp94}).}  
\begin{equation}
\label{anti}
\braket{n_{k,\alpha}}{\theta_{k,\alpha}}_{a} =
\prod_k\frac{k^{r_k/2}}{(2 \pi)^{r_k/2} r_k!} \sum_{\pi_k \in S_{r_k}} 
(-1)^{\sum_k |\pi_k|}\,\esp{- \ii \sum_{k,\alpha} k 
n_{k,\alpha}\theta_{k,\pi_k(\alpha)} }~.
\end{equation}
The integration over the angles $\theta_{k,\alpha}$ gives as a result a set of  
$\delta$-functions
in the momenta $n_{k,\alpha}$, whose structure in related to the cycles of
$\pi_k$, since it forces the momenta associated to the same cycle
to coincide. In the end for each cycle of order $h$ of $\pi_k$ we have one 
integer
momentum and the corresponding partition function is the one of QED on a
torus of area $hkt$, in complete agreement with the discussion of Section 3.
The combinatorial factors are also easily checked. Let $s_h(k)$ be the number
of cycles of order $h$ in $\pi_k$, with $\sum_{h=1}^{r_k} h s_h(k) = r_k$; then 
the total number
of permutations in $\pi_k$ with a given cycle decomposition is 
$r_k!/(\prod_{h=1}^{r_k}
s_h(k)! h^{s_h(k)})$. By inserting this degeneracy into (\ref{extra5}) and 
(\ref{extra4}) we find for the bosonic case
\begin{eqnarray}
 \label{nquattro}
 Z_N^{\rm b}(t) & = &  
 \sum_{\{r_k\}} \delta\Bigl(\sum_{k=1}^{N} k r_k-N\Bigr)
 \sum_{\{s_h(k)\}} \delta\Bigl(\sum_{h=1}^{r_k} h s_h(k)-r_k\Bigr)
 \nonumber\\
 & & 
 \times\prod_{k=1}^N \prod_{h=1}^{r_k}  
 {\left[Z_{\rm QED}(hkt)\right]^{s_h(k)}\over
 s_h(k)! h^{s_h(k)}}~.
\end{eqnarray}
In the fermionic case, besides a factor $(-1)^F$ with 
$F=\sum_k r_k=\sum_{h,k} h s_h(k)$
 one has to introduce also a factor
$(-1)^{\sum_k |\pi_k|} = (-1)^{ \sum_{h,k} (h-1) s_h(k)}$, 
due to the antisymmetrization of the wave functions.
Combining these two signs we have simply to insert
$(-)^{\sum_{h=1}^{r_k} s_h(k)}$. 
We have then
\begin{eqnarray}
 \label{ncinque}
 Z_N^{\rm f}(t) & = &  
 \sum_{\{r_k\}} \delta\Bigl(\sum_{k=1}^{N} k r_k-N\Bigr)
 \sum_{\{s_h(k)\}} \delta\Bigl(\sum_{h=1}^{r_k} h s_h(k)-r_k\Bigr)
 \nonumber\\
 & & 
 \times\prod_{k=1}^N \prod_{h=1}^{r_k}  
 {\left[-Z_{\rm QED}(hkt)\right]^{s_h(k)}\over
 s_h(k)! h^{s_h(k)}}~.
\end{eqnarray}

The grand-canonical partition function is obtained by inserting
(\ref{nquattro}) or (\ref{ncinque}) into
\begin{equation}
 Z^{\rm b,f}(t,q)=\sum_N Z_N^{\rm b,f}(t) q^N~.
\end{equation}
The sum over $s_h(k)$, now unconstrained, can be performed and the result
of the previous section is easily reproduced:
\begin{eqnarray}
 \label{dodici}
  Z^{\rm b,f}(t,q) & = &  
  \exp\left(\mp\sum_k\sum_n \log (1 - q^k {\rm e}^{-2\pi^2 k t n^2})\right)
  \nonumber\\
  & = & \prod_{kn}\left(1-q^k e^{- 2\pi^2 k t n^2}\right)^{\mp 1}~.
\end{eqnarray}    
 
\section{Concluding remarks} 
The  analysis developed in the previous sections led us to a rather surprising
conclusion:
quantization of YM2 on a torus by using the unitary gauge and preserving all
classical symmetries defines  a theory that has a richer  structure 
than the one obtained so far  in the literature by using various gauges or  
lattice regularization.
In the conventional formulation the partition function on the torus has been 
known for some time \cite{r90,bt92}
and it has been given an interpretation in terms of $N$ free fermions
\cite{mp,cdmp94,douglas} on a circle. The corresponding grand canonical
ensemble coincides with the restriction to $k=1$ of the grand canonical 
partition function  we obtain and that is given in (\ref{dodici}). 
The new states with $k>1$ introduced by our analysis are related to non 
trivial permutations $P$ in the compact space direction and they are in one to
one correspondence with the cycles of length $k$ in $P$ in complete analogy 
with the states introduced by \cite{dvv} in the context of Matrix String Theory.

Naturally one would like to reproduce the same results in different 
gauges and understand, for instance, how the new $k>1$ states appear in 
the gauge $\partial_1A_1=0$ with $A_1$ diagonal. 
We do not have yet the full answer to this problems but we can point
out some clues, and in one case a positive evidence, that the new set of
states are required if one wants to preserve modular invariance, or in general
invariance under discrete diffeomorphisms, in the quantization.

The first clue consists in the fact that the restriction to the
$k=1$ states corresponds, as we have seen, to a truncation of the full theory 
to one where the sum over all pairs of commuting permutations $(P,Q)$ 
is replaced by the sum over the subset of  pairs of the
form $({\bf 1},Q)$, which is clearly not a modular invariant subset. 

On the other hand it would not be surprising if the gauge choice 
$\partial_1 A_1=0$, which unlike the unitary gauge is not manifestly modular
invariant, turned out not to be the most convenient  to reveal topological 
structures linked to non trivial permutations on one of the cycles of the torus. 
In fact it is not even granted that $\partial_1 A_1=0$ is  admissible, in the
sense that it might project onto the trivial
topological sector in one of the cycles of the torus\footnote{An example of this 
type is the gauge choice $A_1=0$, which is not admissible if the 
$x$ direction is compactified.}.

It is useful at this point to remember that the $(P,Q)$ sectors are related to
the topological obstructions to a global smooth diagonalization of $F$ on the
torus. There are other topological obstructions of the same type in the theory.
Consider a Wilson loop that winds once around a cycle of the torus:
\eq
W_1(t)~=~P~\exp\left\{\ii\int_0^{2\pi} dx A_1(x,t)\right\}~.
\label{m1}
\en
If the loop is moved once around the other cycle of the torus, its eigenvalues
will in general undergo a permutation $Q$:
\eq
\diag~ W_1(t+2\pi) ~=~ Q^{-1} \diag~W_1(t)~Q~.
\label{m2}
\en
The same argument obviously applies to $W_0(x)$, when $x$ is increased of
$2\pi$:
\eq
\diag~ W_0(x+2\pi) ~=~ P^{-1} \diag~W_0(x)~P ~.
\label{m3}
\en
However  $W_1(t)$ and $W_0(x)$ in general do not commute  and they cannot be
diagonalized simultaneously.  

Consider now the theory on a cylinder with the dimension $x$ compactified and
the edges in correspondence with $t=0$ and $t=2\pi$ . The partition function of
the torus is obtained by identifying $\diag\, W_1(0)$ and  $\diag\, W_1(2\pi)$
up to a gauge transformation, namely, in the gauge  $\partial_1 A_1=0$ with
$A_1$ diagonal, by identifying $\diag\, W_1(0)$ and  $\diag\, W_1(2\pi)$
up to an arbitrary permutation $Q$ of the eigenvalues. What about the sectors
corresponding to the non trivial permutation $P$ of the eigenvalues of $W_0(x)$
(Eq.(\ref{m3}))? 
Have they been taken into account automatically by sewing the two
ends of the cylinder with a group integration and a sum over all permutations
$Q$?  According to our discussion in the unitary gauge the answer to the last
question is no.
In fact it has been shown in the last section that in a sector where
\eq
\diag\, F(x+2\pi,t) = P^{-1} \diag\, F(x,t) P 
\en
the independent eigenvalues of $W_1(0)$ and $W_1(2\pi)$ are in one to one
correspondence with the cycles of $P$.  More precisely the eigenvalues of $W_1$
corresponding to a cycle of $P$ of length $k$ are of the form
\eq
\esp{\ii\frac{\phi}{k}+\frac{2\ii\pi r}{k}}~~ (r=0,\cdots,k-1)~.
\label{m4}
\en
When sewing the ends of the cylinder, only eigenvalues corresponding to cycles
of the same length can be identified, which is tantamount to restrict the
permutation $Q$ to commute with $P$.

So in the unitary gauge the standard integration over the group manifold 
parametrized by
the invariant angles of the holonomy projects onto the trivial sector $P={\bf 1}$ 
in the compactified direction of the cylinder.
The sum over a complete set of states requires instead to consider a permutation 
$P$ and decompose it into cycles.
Let $r_k$ be the number of cycles of length $k$ and $\phi_\alpha^{(k)}
(\alpha=1,\cdots,r_k)$ the invariant angles associated to each
cycle\footnote{This is the eigenvalue of a Wilson loop that winds $k$ times
around the cylinder. All its eigenvalues associated to a cycle of length $k$
then coincide and the corresponding invariant angles are periodic of period
$2\pi$.}.  The identification of the eigenvalues when sewing the ends of the
cylinder is done modulo permutations of the angles corresponding to the same
length $k$ and the integration volume is not the one of U$(N)$ but rather of
${\rm U}(r_1)\otimes {\rm U}(r_2)\otimes {\rm U}(r_3)\cdots$.

It would be desirable to show that the same complete set of states is required
in other gauges if one wants to include the configurations corresponding to a
non trivial $P$ in Eq.~(\ref{m3}).
This is technically not easy because $W_1$ and $W_0$ do not commute and the
corresponding functional integrals are more involved.
There is however one case in which $W_0$ and $W_1$ commute, namely the model
with $t=0$ (the BF theory) where the functional integral over $F$ leads to a
$\delta[f(A)]$. This case is studied in Appendix C where it is shown that by
a suitable gauge transformation
\eq
W_1 \rightarrow \diag\, W_1(t)\,P~, \hskip 1cm 
W_0 \rightarrow \diag\, W_0(x)\,Q~,
\en
and that the eigenvalues of $\diag\, W_1(t)$ (resp.  $\diag\, W_0(x)$) 
corresponding to the same cycle of $P$ (resp. $Q$) coincide.
Moreover Eq.s (\ref{m2}) and (\ref{m3}) hold and $[P,Q]=0$. The eigenvalues of
$\diag\,W_1(t)\,P$ and of $\diag\,W_0(x)\,Q$ 
then follow the pattern of Eq. (\ref{m4})
and the sum over a complete set of states is done accordingly to the
prescription discussed above.

Although $t=0$ is a singular point, where the partition function becomes the
volume of the moduli space of a flat connection, it is nevertheless important
that the results obtained in the unitary gauge scheme are consistently
reproduced in this case  by diagonalizing the non contractable Wilson loops.

We remarked earlier about the close analogy between the spectrum obtained here
and the states in Matrix string theory described in~\cite{dvv}. Moreover, just
as in Matrix string theory, in the $t\to\infty$ limit the states with $n_i>0$ 
decouple and we are left with the partition function of a 
conformal field theory. 

In the present framework the states of the spectrum do not interact. The
interaction may be implemented by allowing the gauge fixing matrix $M(\tau,x)$
introduced in appendix B to have branched points. Take for instance a square
root branch point at $\tau=0$ involving two eigenvalues $\lambda_i$ and
$\lambda_j$. The two eigenvalues do not cross each other for $\tau<0$, but they do
for $\tau>0$. So if $i$ and $j$ are contained in the same cycle (string) at $\tau<0$
the cycle (string) will break into two for $\tau>0$ (similar mechanisms are
discussed in~\cite{dvv,bbn,bbn2}). In general string interaction will be 
described by configurations where the eigenvalues live on higher genus 
Riemann surfaces which are branched coverings of the original torus. A
different, although possibly related,  problem is
how to quantize YM2 on a surface with non vanishing curvature, namely how to
consistently regularize the divergences appearing in Eq. (\ref{anomaly}),
while preserving
the structure discovered in the case of the torus and invariance under
discrete diffeomorphisms (which might amount to the same thing). This is an
open problem whose solution, if it exists, might require a functional
integration over all metrics, namely quantization of 2d gravity itself
or alternatively a supersymmetric extension of the model.

Another open question concerns the relevance of the states with $k>1$ to the
large $N$ limit, and in particular to the interpretation of YM2 as a string
theory given by Gross and Taylor~\cite{gt}.
We remark  that the large $N$ limit originally 
introduced by 't Hooft~\cite{tH} and considered in  ~\cite{gt} corresponds to
scale $t$ with $N$ according to  $t=\tilde t/N$ with constant
$\tilde t$. In the large $N$ limit $t$ goes to zero and the partition function is 
dominated by the contribution from small cycles as discussed in Section 3.
On the contrary the long string states, analogue to the Matrix string states of 
\cite{dvv}, are the leading contributions at large $t$.
Furthermore the scaling of $t$ with $N$ is not compatible with the grand 
canonical partition function formulation, 
which is the natural framework to describe the states of
arbitrary length and requires summing over all $N$ at fixed $t$.
All these considerations point to the fact that the string picture emerging from
the analogy with the Matrix string theory is distinct from the one of Gross
and Taylor, although it is possible that the two pictures are related by some
strong-weak coupling duality\footnote{For this duality to be apparent in our formulas
it would be necessary to include string interaction to all orders.}.
The grand canonical formulation contains a new parameter
$\mu$, the chemical potential.
This is reminiscent of the IKKT matrix model~\cite{IKKT} where a sum over all 
matrix sizes
is required to make contact with superstring theory. 
In conclusion the quantization of 2d  Yang-Mills theory with $U(N)$ gauge group
seems naturally to lead to some more general underlying 
theory. It is possible then that the analogy with Matrix string theory of 
~\cite{dvv} is more than just a formal analogy, and that a deeper understanding of the stringy nature of 
2d Yang-Mills theory may provide us with a deeper 
insight of Matrix theory as well.

\vskip 1cm
{\bf Note added}
\renewcommand{\theequation}{\alph{equation}}
\setcounter{equation}{0}
\par\noindent
Immediately after the first version of this paper was submitted to hep-th, 
the partition function of the DVV model in the IR
limit was computed in \cite{kv} where the possibility that the computation 
might give the exact result was also pointed out. 
The result (eq. 2 in \cite{kv}) coincides {\em exactly}
with the logarithm of our ``bosonic'' partition function $Z_N^{\rm b}$
at fixed $N$ (see eq. (\ref{gc2}):
\begin{equation}
\label{note1}
Z_N^{\rm DVV}(t) = \log Z_N^{\rm b}(t) = \sum_{hk = N} {1\over h} 
\sum_n \esp{-2\pi^2 hktn^2} = \sum_{h|N} \esp{-2\pi^2 Ntn^2}~.
\end{equation}
This nicely substantiates our concluding remark that the relation between 
U$(N)$ YM theory on a cylinder or torus and Matrix Strings is more than a
formal analogy. The result (\ref{note1}) arises because the matter fields $X_i$
and $\psi_\alpha$ of the DVV model do not contribute, because of  supersymmetry, 
to the partition function (see eq. (30) in \cite{kv}) which is then due entirely 
to the U$(N)$ gauge field.
However, the structure of fermionic 0-modes is argued to effectively kill the
``disconnected'' contributions, i.e. those arising from configurations which
in our language have more
than one ``connected block'' of eigenvalues; this 
clearly accounts for the logarithm.  

\vskip 1cm
{\bf  Acknowledgements}
\par\noindent
Work supported in part by the European Commission TMR programme 
ERBFMRX-CT96-0045.
\newpage

\appendix{\Large {\bf{ A. Commuting permutations}}}
\vskip 0.5cm
\renewcommand{\theequation}{A.\arabic{equation}}
\setcounter{equation}{0}
The topological sectors described above are  labelled by ordered pairs 
of commuting permutations. Therefore we need an explicit construction 
of all the permutations of $N$ elements commuting with a given 
permutation $P$.
\par
Let $r_k,\ (k=1,\dots,N)$, with $\sum_k r_k=N$, 
be the number of cycles of length $k$ in the 
permutation $P$. 
Let us denote the elements of the
set $\{1,\dots,N\}$ with a three--index notation based on how they transform
under P:
\be
a^{k,\alpha}_n \qquad (k=1,\dots,N)\ (\alpha=1,\dots,r_k)\ (n=1,\dots,k)
\ee
is the $n$--th elements in the $\alpha$--th cycle of length $k$.
Therefore 
\be
P(a^{k,\alpha}_n)=a^{k,\alpha}_{n+1}~,
\ee
where $n+1$ is understood ${\rm mod}\ k$.
\par
Let $Q$ be a permutation commuting with $P$ and consider its action on 
the cycle $(a^{k,\alpha}_1,\dots,a^{k,\alpha}_k)$ of $P$. We have
\be
QP(a^{k,\alpha}_n)=Q(a^{k,\alpha}_{n+1})
\ee
and therefore, using $PQ=QP$,
\be
P\left(Q(a^{k,\alpha}_n)\right)=Q(a^{k,\alpha}_{n+1})~,
\ee
which means that $\left(Q(a^{k,\alpha}_1),\dots,Q(a^{k,\alpha}_k)\right)$  
is a cycle of length $k$ in $P$. Hence, there exists a permutation 
$\pi_k\in S_{r_k}$ of $r_k$ elements such that the following equality between 
cycles holds
\be
\left(Q(a^{k,\alpha}_1),\dots, Q(a^{k,\alpha}_k)\right)=
\left(a^{k,\pi_k(\alpha)}_1,\dots,a^{k,\pi_k(\alpha)}_n\right)~.
\ee
This implies that there exist $r_k$ integers 
\be
s(k,\alpha)\ 
(\alpha=1,\dots,r_k)\ (1\le s(k,\alpha)\le k)
\ee 
such that
\be
Q(a^{k,\alpha}_n)=a^{k,\pi_k(\alpha)}_{n+s(k,\alpha)}\label{q}~,
\ee
where $n+s(k,\alpha)$ is understood ${\rm mod}\ k$.
\par
We have thus shown that a permutation $Q$, 
commuting with $P$, is completely determined by
assigning for each $k=1,\dots,N$
\begin{itemize}
\item a permutation $\pi_k\in S_{r_k}$, where $r_k$ is the number of 
cycles of length $k$ in $P$;
\item a set of $r_k$ integers $1\le s(k,\alpha)\le k$.
\end{itemize}
The permutation $Q$ is then defined by Eq.~(\ref{q}). 
This shows in particular that the number
of permutations $Q$, commuting with $P$, is 
\be
|C(P)|=\prod_k r_k! k^{r_k}~.
\ee

\vskip .8 cm
\appendix{\Large {\bf{ B. BRST formalism}}}
\vskip 0.5cm
\renewcommand{\theequation}{B.\arabic{equation}}
\setcounter{equation}{0}
We develop in this appendix the BRST formalism for YM2 on a torus in the Unitary
gauge and discuss how the non trivial sectors considered in 
Section \ref{sec3} arise in this context.
The first order action introduced in (\ref{partriemann}) is invariant under
gauge transformations
\begin{eqnarray}
\delta A& = &d\epsilon - \ii [A,\epsilon]~, \nonumber \\
\delta F& = & -\ii [F,\epsilon]~. 
\label{gaugeinv}
\end{eqnarray}
Correspondingly, the BRST  and anti-BRST transformations are given by
\begin{eqnarray}
&s A = d c - \ii [A,c]~;~~~~~~~~~~&\bar{s} A= d \bar{c} -\ii [A,\bar{c}]~,
 \nonumber \\
&s F =  -\ii [F,c]~;~~~~~~~~~~~~~~&\bar{s} F=  -\ii [F,\bar{c}]~, \nonumber \\
&s c= \ii c c~;~~~~~~~~~~~~~~~~~~~~~&\bar{s} \bar{c}=\ii \bar{c} \bar{c}~, 
\nonumber \\
&s \bar{c}= \ii c \bar{c} + b~;~~~~~~~~~~~~~~~~&\bar{s} c= \ii \bar{c} c -b~,
\nonumber \\
&s b= \ii c b~;~~~~~~~~~~~~~~~~~~~~~&\bar{s} b= \ii \bar{c}b~, 
\label{brst}
\end{eqnarray}
where all fields are hermitian $N \times N$ matrices.
In order to fix the gauge let us introduce a matrix $M(\tau,x)$ and add to the 
action a BRST and anti-BRST invariant term of the type
\be
S_{\rm g.f} = \int_0^{2 \pi} d\tau dx\, \tr~ s\bar{s} (M F)
\label{gfix}
\ee
which, using the BRST transformations (\ref{brst}), takes the form
\be
S_{\rm g.f} = \int_0^{2 \pi} d\tau dx~ \tr
\left( M c F \bar{c} - M \bar{c} F c +
M \bar{c} c F -M c  \bar{c} F + \ii b [F,M] \right)~.
\label{gfix2}
\ee
The functional integration over the auxiliary field $b$ leads to a
$\delta$-function of argument $[F,M]$, which implies that in the base of the
eigenvectors of $M$ the matrix $F$ is diagonal and the unitary gauge is
implemented. 
It is convenient therefore to rewrite
Eq. (\ref{gfix2}) in the base of the eigenvectors of the gauge-fixing matrix 
$M$. As already discussed in Section \ref{sec3}, 
when the non trivial sectors were
introduced, matrices on a torus can be divided into classes characterised by a 
pair of commuting permutations $(P,Q)$. They are  the permutations of the
eigenvalues obtained if we go round the non contractable loops $(a,b)$ that
generate the fundamental group of the torus\footnote{Clearly the assumption of
continuity  $M$ and of its first derivatives must be made here.}.
It easy to see that a gauge fixing matrix $M$ belonging to the class 
$(P,Q)$ defines a functional integral over field configurations of the sector
$(P,Q)$ defined in Section \ref{sec3}.
In fact if we denote by $\alpha_i(\tau,x)$ the eigenvalues of $M$, we have
\be
\alpha_i(\tau+2 \pi,x) = \alpha_{P(i)}(\tau,x)~,~~~~~~\alpha_i(\tau,x+2 \pi) =
\alpha_{Q(i)}(\tau,x)~,
\label{bc}
\ee
and the same boundary conditions are obeyed by all the other field in the base
of the eigenvectors of $M$. 
In this base the BRST invariant action can be explicitly written as
\begin{eqnarray}
S_{\rm BRST}&=& \int_0^{2 \pi} d\tau dx   \sum_{i,j} 
\Bigl[\ii b_{ij} (\alpha_i -\alpha_j) F_{ji} + (\lambda_i-\lambda_j)(\alpha_i 
-\alpha_j)
\bar{c}_{ij} c_{ji} +\nonumber \\
& + & (\lambda_i-\lambda_j) A_{0,ij}A_{1,ji}\Bigr]\nonumber\\   
& + &\int_0^{2 \pi} d\tau dx   \sum_i \Bigl[ \frac{t}{2} 
\lambda_i^2 - \ii \lambda_i (\partial_0 A_1^{(i)} -\partial_1 A_0^{(i)} )~, 
\Bigr]
\label{brstinv}
\end{eqnarray}
where the diagonal elements of $F$ have been denoted $\lambda_i$ and the
diagonal elements of $A_{\mu}$ by $A_{\mu}^{(i)}$.
Besides, as already mentioned, all fields appearing in (\ref{brstinv}) satisfy 
the same boundary condition (\ref{bc}) as $\alpha_i$. 
It is clear that the unitary gauge condition $F_{ji}=0$ for $i \neq j$ is
implemented by the functional integral over $b_{ij}$ and BRST invariance 
ensures that the dependence from the eigenvalues $\alpha_i$ of the gauge fixing
matrix $M$ cancel, as it can be seen by performing explicitly the functional 
integration over both $b_{ij}$ and the ghost anti-ghost fields. 
The supersymmetry
(\ref{supersy}), suitably modified\footnote{The factor $(\alpha_i -\alpha_j)$ in
the term containing the ghost fields should be absorbed by redefining $c$ and
$\bar{c}$ to reproduce (\ref{supersy}).}, on the other hand, ensures that the 
Vandermonde determinants $\Delta(\lambda)$ coming from the integration over the
ghost system are exactly canceled by the result of the integration over the non
diagonal part of $A_{\mu}$.

Finally it should be noticed that the gauge has not been completely fixed in
(\ref{brstinv}), the action being still invariant under a local U$(1)$ symmetry
for each eigenvalue $\lambda_i$. Correspondingly the diagonal part of $c$, 
$\bar{c}$ and $b$ do not appear in (\ref{brstinv}).

In conclusion the sectors described in Section \ref{sec3} are generated by gauge 
fixing
condition which are not connected by smooth variations of the the gauge fixing
matrix $M$, hence BRST invariance does not fix the relative weight of the
different sectors in the partition function.

\vskip .8 cm
\appendix{\Large {\bf{ C. Topological obstructions in the BF model}}}
\vskip 0.5cm
\renewcommand{\theequation}{C.\arabic{equation}}
\setcounter{equation}{0}
Let us consider the action (\ref{partriemann}) with $t=0$. The functional 
integral over $F$ produces a $\delta(f(A))$ which has the solution:
\begin{equation}
A_{\mu}(x,\tau) = \ii g^{-1}(x,\tau) \partial_{\mu} g(x,\tau)~. 
\label{puregauge}
\end{equation}
Consider now the non contractable Wilson loops
\begin{eqnarray}
W_0(x,\tau)& = &P \exp \left\{\ii \int_{\tau}^{\tau+2 \pi} A_0(x,t) dt\right\}
= g^{-1}(x,\tau)g(x,\tau+2 \pi)~, \nonumber \\
W_1(x,\tau)& = &P \exp \left\{\ii \int_{x}^{x+2 \pi} A_1(y,\tau) dy \right\}
= g^{-1}(x,\tau)g(x+2 \pi,\tau)~. 
\label{Wil}
\end{eqnarray}
As $A_{\mu}(x,\tau)$ are defined globally on the torus, namely they are periodic
in both variables, it follows from (\ref{Wil}) that $W_0(x,\tau)$ and 
$W_1(x,\tau)$
are also periodic in both $x$ and $\tau$.
By using the explicit form of $W_0(x,\tau)$ and $W_1(x,\tau)$ in terms of 
$g(x,\tau)$
one easily finds
\begin{equation}
[W_0(x,\tau),W_1(x,\tau)]=0~.
\label{commutano}
\end{equation}
Notice also that from the definition above we have
\begin{eqnarray}
g(x,\tau+2 \pi)& =& g(x,\tau) W_0(x,\tau)~, \nonumber\\
g(x+2 \pi,\tau)& =&  g(x,\tau) W_1(x,\tau)~. \label{gshift}
\end{eqnarray}
Due to Eq.(\ref{commutano}) it is always possible to find {\it locally}
a unitary transformation $U(x,\tau)$ that diagonalizes both $W_0(x,\tau)$
and $W_1(x,\tau)$:
\begin{eqnarray}
W_0(x,\tau)& =& U^{-1}(x,\tau) w_0(x) U(x,\tau)~, \nonumber \\
W_1(x,\tau)& =& U^{-1}(x,\tau) w_1(\tau) U(x,\tau)~, 
\label{diag}
\end{eqnarray}
where $w_0(x)$ and $w_1(\tau)$ are the diagonal matrices displaying the
eigenvalues of $W_0(x,\tau)$ and $W_1(x,\tau)$. Notice that the eigenvalues
of $W_0(x,\tau)$ are only function of $x$ and the ones of $W_1(x,\tau)$ only
of $\tau$; in fact it follows from the definition (\ref{Wil}) that for instance
$W_0(x,\tau')$  is related to $W_0(x,\tau)$ by a unitary transformation.
There are in general topological obstructions to a global and smooth 
diagonalization of $W_0(x,\tau)$ and $W_1(x,\tau)$; as a result 
$w_0(x+ 2 \pi)$ and $w_1(\tau +2 \pi)$ will coincide with
$w_0(x)$ and $w_1(\tau)$ only up to an element of the Weyl group,
namely, for U$(N)$, up to a permutation:
\begin{equation}
 w_0(x+ 2 \pi)=P w_0(x)P^{-1},~~~~~~~~~~~w_1(\tau +2 \pi)=Qw_1(\tau)Q^{-1}~.
\label{shiftw}
\end{equation}
Due to the periodicity of  $W_0(x,\tau)$ and $W_1(x,\tau)$ we must have
also
\begin{equation}
U(x+2 \pi,\tau)= P  U(x,\tau)~,~~~~~~~~~~~U(x,\tau+2 \pi)= Q  U(x,\tau)~,
\label{shiftU}
\end{equation}
which entails
\begin{equation}
[P,Q]=0~.
\label{commPQ}
\end{equation}
It follows from (\ref{diag}) that the eigenvalues of $w_0(x) $ 
and $ w_1(\tau)$ are not all independent. 
In fact if we shift $\tau$ (resp. $x$) by $2 \pi$ in the first 
(resp. second) of Eq.s (\ref{diag}) we obtain from Eq. (\ref{shiftw}) and the 
periodicity of $W_0(x,\tau)$ (resp $W_1(x,\tau)$ ) the following constraints:
\begin{equation}
w_1(\tau)=P^{-1} w_1(\tau) P~;~~~~~~~~~~~~~~ w_0(x)=Q^{-1} w_0(x) Q~.
\label{consw}
\end{equation}
These conditions are satisfied if all the eigenvalues of $w_1(\tau)$ 
which are  mapped into each other by $P$ coincide; in other words, 
the eigenvalues of $w_1(\tau) $ are associated to the cycles of $P$. 
The same applies to $w_0(x)$ and $Q$.
We want to remark at this point that the unitary transformation 
(\ref{diag}) is {\it not} a gauge transformation. 
In fact the correct gauge transformation of $W_0(x,\tau)$ and $W_1(x,\tau)$
with the unitary matrix $U(x,\tau)$ is given by
\begin{eqnarray}
U(x,\tau) W_0(x,\tau) U^{-1}(x,\tau+2 \pi)& =& w_0(x) Q^{-1}~, \nonumber \\
U(x,\tau) W_1(x,\tau) U^{-1}(x+2 \pi,\tau)& =& w_1(\tau) P^{-1}~. 
\label{gaugetrW}
\end{eqnarray}
The correct gauge transformed Wilson loops are then $ w_0(x) Q^{-1}$ and 
$w_1(\tau) P^{-1}$ rather than $ w_0(x) $ and $w_1(\tau)$. 
If we diagonalize one of them,
say $ w_0(x) Q^{-1}$, by a constant gauge transformation\footnote{The constant 
unitary transformation that diagonalizes each block in $Q$ \
corresponding to a cycle does not affect the diagonal form of $w_0(x) $.} 
then we may conclude that the eigenvalues 
of the gauge transformed Wilson loop have the form given in (\ref{m4}) for each 
cycle of $Q$.
The implications of this can be better understood from the following example. 
Consider the theory on a cylinder with compactified dimensions $x$ and 
$\tau$ ranging between $0$ and $2 \pi$.
Suppose we identify the states with the holonomies on the boundaries, 
namely with the configurations of $w_1(\tau)$ that satisfy the first of 
Eq.s (\ref{consw}) with $P=1$. We can obtain the torus
by sawing the two ends of the cylinder, that is by identifying the 
holonomies at the ends up to a permutation $Q$ of the eigenvalues 
according to the second of Eq.s (\ref{shiftw}).
The sum over $Q$ implies that in the channel obtained by cutting the torus at 
constant $x$ we recover the whole spectrum of states satisfying 
the constraints (\ref{consw}) with arbitrary $Q$.


\begin{thebibliography}{99}

\bibitem{m75} A.A. Migdal, Sov. Phys. JETP {\bf 42}, (1975) 413. 

\bibitem{r90} B. Ye. Rusakov, \MPL{A5} (1990) 693.

\bibitem{w91} E. Witten, Comm.. Math. Phys. {\bf 141} (1991) 153. 

\bibitem{w92} E. Witten, J. Geom. Phys {\bf 9} (1992) 303. 

\bibitem{bt92} M. Blau and G. Thompson, \IJMP{A7} (1992) 3781.

\bibitem{blth} M. Blau and G. Thompson,{\it ``Lectures on 
$2d$ Gauge Theories - Topological Aspects and Path Integral Techniques"}, 
Proc. 1993 Summer School in High Energy Physics and Cosmology,
ed.s E. Gava et al. (Trieste 1993), World Scientific,1994,
\hepth{9310144} 
\bibitem{gpss}
G. Grignani, L. Paniak, G.W. Semenoff and P. Sodano,
Annals Phys {\bf 260} (1997) 275.

\bibitem{dk} M.R. Douglas and V.A. Kazakov, \PL{B319} (1993) 219.

\bibitem{cdmp93} M. Caselle, A. D'Adda, L. Magnea and S. Panzeri,
Proc. 1993 Trieste Summer
School in High Energy Physics and Cosmology (Trieste 1993), 
ed.s E. Gava et al., World Scientific, 1994, \hepth{9309107}.

\bibitem{gm1} D.J. Gross and A. Matytsin, \NP{B429} (1994) 50.

\bibitem{gm2} D.J. Gross and A. Matytsin, \NP{B437} (1994) 541.

\bibitem{bg} A. Bassetto and L. Griguolo, {\it Two-dimensional QCD, 
instanton contributions and the perturbative Wu-Mandelstam-Leibbrandt
prescription}, \hepth{9806037}.

\bibitem{tH} G. 't Hooft, \NP{B75} (1974) 461.

\bibitem{gt} D.J. Gross, \NP{B400} (1993) 161;
D. Gross and W. Taylor, \NP{B400} (1993) 181 and \NP{B403} (1993) 395.

\bibitem{motl} L. Motl, {\it Proposals on nonperturbative superstring 
interactions}, \hepth{9701025}.

\bibitem{dvv} R. Dijkgraaf, E. Verlinde and H. Verlinde, Nucl.Phys. B500 (1997)
43.
 
\bibitem{bc} L. Bonora and C.S. Chu, \PL{B410} (1997) 142.

\bibitem{bfss} T. Banks, W. Fischler, S. H. Shenker and L. Susskind, 
Phys. Rev. {\bf D55} (1997) 5112.

\bibitem{wt} Washington Taylor IV, \PL{B394} (1997) 283.

\bibitem{IKKT} N. Ishibashi, H. Kawai, Y. Kitazawa and A. Tsuchiya, 
\NP{B498} (1996) 467.

\bibitem{h94} J.E. Hetrick, \IJMP{A9} (1994) 3153.

\bibitem{bt94}  M. Blau and G. Thompson, \CMP{171} (1995) 639.

\bibitem{cdmp94} M. Caselle, A. D'Adda, L. Magnea and S. Panzeri, \NP{B416}
(1994) 751.

\bibitem{peri} V. Periwal and O. Tafjord, 
{\it Mechanism for long Dijkgraaf-Verlinde-Verlinde strings},
\hepth{9710204}.

\bibitem{mp} J.A. Minahan and A.P. Polychronakos, \NP{B422}, (1994) 172 

\bibitem{douglas} M. Douglas, 
{\it Conformal field theory techniques for large N group theory},
\hepth{9303159}.

\bibitem{bbn}
G. Bonelli, L. Bonora and F. Nesti, 
{\em String interactions from matrix string theory},
\hepth{9805071}.

\bibitem{bbn2} G. Bonelli, L. Bonora and F. Nesti, 
{\it Matrix string theory, 2-d SYM instantons and affine Toda systems},
\hepth{9807232}.

 
\bibitem{kv} I.K. Kostov and P. Vanhove, 
{\it Matrix String partition functions}, \hepth{9809130}.
 

\end{thebibliography}
\end{document}